\documentclass[twocolumn]{aastex631}
\usepackage{CJKutf8} 
\usepackage{amsmath}
\usepackage{comment}
\usepackage{physics}
\usepackage{verbatim}

\usepackage{soul}

\begin{document}

\title{Signatures of a Tidally Induced Spiral Arm at the Anticenter of the Milky Way and a Kinematically Extended Anticenter Stream Using DESI DR2}
  
\correspondingauthor{Mika Lambert}
\email{mlamber1@ucsc.edu}

\author[0000-0002-2527-8899]{Mika Lambert}
\affiliation{Department of Astronomy \& Astrophysics, University of California, Santa Cruz, 1156 High Street, Santa Cruz, CA 95064, USA}

\author[0000-0002-6667-7028]{Constance M. Rockosi}
\affiliation{Department of Astronomy \& Astrophysics, University of California, Santa Cruz, 1156 High Street, Santa Cruz, CA 95064, USA}

\author[0000-0003-2644-135X]{Sergey E. Koposov}
\affiliation{Institute for Astronomy, University of Edinburgh, Royal Observatory, Blackford Hill, Edinburgh EH9 3HJ, UK}

\author[0000-0002-9110-6163]{Ting S. Li}
\affiliation{Department of Astronomy \& Astrophysics, University of Toronto, Toronto, ON M5S 3H4, Canada}

\author[0000-0002-6257-2341]{Monica Valluri}
\affiliation{Department of Astronomy, University of Michigan, Ann Arbor, MI 48109, USA}
\affiliation{University of Michigan, 500 S. State Street, Ann Arbor, MI 48109, USA}

\author[0000-0002-0740-1507]{Leandro {Beraldo e Silva}}
\affiliation{Observat\'orio Nacional, Rio de Janeiro—RJ, 20921-400, Brasil }

\author[0000-0002-6469-8263]{Songting Li}
\affiliation{Department of Astronomy, School of Physics and Astronomy, and Shanghai Key Laboratory for Particle Physics and Cosmology, Shanghai Jiao Tong University, Shanghai 200240, People's Republic of China}
\affiliation{State Key Laboratory of Dark Matter Physics, School of Physics and Astronomy, Shanghai Jiao Tong University, Shanghai 200240, China}

\author[0000-0002-7662-5475]{Jo\~ao A. S. Amarante}
\affil{Department of Astronomy, School of Physics and Astronomy, \\ Shanghai Jiao Tong University, 800 Dongchuan Road, Shanghai, 200240, China}
\affil{State Key Laboratory of Dark Matter Physics, School of Physics and Astronomy,\\ Shanghai Jiao Tong University, Shanghai, 200240, China}

\author[0000-0002-5689-8791]{Amanda Bystr\"om}
\affil{Institute for Astronomy, University of Edinburgh, Royal Observatory, Blackford Hill, Edinburgh EH9 3HJ, UK}

\author[0000-0003-0105-9576]{Gustavo E. Medina}
\affil{Department of Astronomy \& Astrophysics, University of Toronto, 50 St George Street, Toronto ON M5S 3H4, Canada}

\author[0000-0002-7393-3595]{Nathan R. Sandford}
\affil{Department of Astronomy \& Astrophysics, University of Toronto, 50 Saint George Street, Toronto, ON M5P 0A2, Canada}

\author[0000-0002-5758-150X]{Joan Najita}
\affil{NSF NOIRLab, 950 N. Cherry Ave., Tucson, AZ 85719, USA}

\author{Namitha Kizhuprakkat}
\affil{Institute of Astronomy and Department of Physics, National Tsing Hua University, Hsinchu 30013, Taiwan}
\affil{Center for Informatics and Computation in Astronomy, National Tsing Hua University, Hsinchu 30013, Taiwan}

\author{Jessica N. Aguilar}
\affil{Lawrence Berkeley National Laboratory, 1 Cyclotron Road, Berkeley, CA 94720, USA}

\author[0000-0001-6098-7247]{Steven Ahlen}
\affil{Department of Physics, Boston University, 590 Commonwealth Avenue, Boston, MA 02215 USA}

\author[0000-0001-9712-0006]{Davide Bianchi}
\affil{Dipartimento di Fisica ``Aldo Pontremoli'', Universit\`a degli Studi di Milano, Via Celoria 16, I-20133 Milano, Italy}
\affil{INAF-Osservatorio Astronomico di Brera, Via Brera 28, 20122 Milano, Italy}

\author[0000-0001-6098-7247]{David Brooks}
\affil{Department of Physics \& Astronomy, University College London, Gower Street, London, WC1E 6BT, UK}

\author{Todd Claybaugh}
\affil{Lawrence Berkeley National Laboratory, 1 Cyclotron Road, Berkeley, CA 94720, USA}

\author[0000-0002-0553-3805]{Kyle Dawson}
\affil{Department of Physics and Astronomy, The University of Utah, 115 South 1400 East, Salt Lake City, UT 84112, USA}

\author[0000-0002-1769-1640]{Axel de la Macorra}
\affil{Instituto de F\'{\i}sica, Universidad Nacional Aut\'{o}noma de M\'{e}xico,  Circuito de la Investigaci\'{o}n Cient\'{\i}fica, Ciudad Universitaria, Cd. de M\'{e}xico  C.~P.~04510,  M\'{e}xico}

\author{Peter Doel}
\affil{Department of Physics \& Astronomy, University College London, Gower Street, London, WC1E 6BT, UK}

\author[0000-0002-2890-3725]{Jaime E. Forero-Romero}
\affil{Departamento de F\'isica, Universidad de los Andes, Cra. 1 No. 18A-10, Edificio Ip, CP 111711, Bogot\'a, Colombia}
\affil{Observatorio Astron\'omico, Universidad de los Andes, Cra. 1 No. 18A-10, Edificio H, CP 111711 Bogot\'a, Colombia}

\author[0000-0001-9632-0815]{Enrique Gazta\~naga}
\affil{Institut d'Estudis Espacials de Catalunya (IEEC), c/ Esteve Terradas 1, Edifici RDIT, Campus PMT-UPC, 08860 Castelldefels, Spain}
\affil{Institute of Cosmology and Gravitation, University of Portsmouth, Dennis Sciama Building, Portsmouth, PO1 3FX, UK}
\affil{Institute of Space Sciences, ICE-CSIC, Campus UAB, Carrer de Can Magrans s/n, 08913 Bellaterra, Barcelona, Spain}

\author[0000-0003-3142-233X]{Satya Gontcho A Gontcho}
\affil{Lawrence Berkeley National Laboratory, 1 Cyclotron Road, Berkeley, CA 94720, USA}
\affil{University of Virginia, Department of Astronomy, Charlottesville, VA 22904, USA}

\author{Gaston Gutierrez}
\affil{Fermi National Accelerator Laboratory, PO Box 500, Batavia, IL 60510, USA}

\author[0000-0003-0201-5241]{Dick Joyce}
\affil{NSF NOIRLab, 950 N. Cherry Ave., Tucson, AZ 85719, USA}

\author[0000-0001-6356-7424]{Anthony Kremin}
\affil{Lawrence Berkeley National Laboratory, 1 Cyclotron Road, Berkeley, CA 94720, USA}

\author[0000-0002-6731-9329]{Claire Lamman}
\affil{The Ohio State University, Columbus, 43210 OH, USA}

\author[0000-0003-1838-8528]{Martin Landriau}
\affil{Lawrence Berkeley National Laboratory, 1 Cyclotron Road, Berkeley, CA 94720, USA}

\author[0000-0001-7178-8868]{Laurent Le Guillou}
\affil{Sorbonne Universit\'{e}, CNRS/IN2P3, Laboratoire de Physique Nucl\'{e}aire et de Hautes Energies (LPNHE), FR-75005 Paris, France}

\author[0000-0003-4962-8934]{Marc Manera}
\affil{Departament de F\'{i}sica, Serra H\'{u}nter, Universitat Aut\`{o}noma de Barcelona, 08193 Bellaterra (Barcelona), Spain}
\affil{Institut de F\'{i}sica d’Altes Energies (IFAE), The Barcelona Institute of Science and Technology, Edifici Cn, Campus UAB, 08193, Bellaterra (Barcelona), Spain}

\author[0000-0002-1125-7384]{Aaron Meisner}
\affil{NSF NOIRLab, 950 N. Cherry Ave., Tucson, AZ 85719, USA}

\author{Ramon Miquel}
\affil{Instituci\'{o} Catalana de Recerca i Estudis Avan\c{c}ats, Passeig de Llu\'{\i}s Companys, 23, 08010 Barcelona, Spain}
\affil{Institut de F\'{i}sica d’Altes Energies (IFAE), The Barcelona Institute of Science and Technology, Edifici Cn, Campus UAB, 08193, Bellaterra (Barcelona), Spain}

\author[0000-0002-2733-4559]{John Moustakas}
\affil{Department of Physics and Astronomy, Siena University, 515 Loudon Road, Loudonville, NY 12211, USA}

\author{Adam Myers}
\affil{Department of Physics \& Astronomy, University  of Wyoming, 1000 E. University, Dept.~3905, Laramie, WY 82071, USA}

\author[0000-0001-9070-3102]{Seshadri Nadathur}
\affil{Institute of Cosmology and Gravitation, University of Portsmouth, Dennis Sciama Building, Portsmouth, PO1 3FX, UK}

\author[0000-0002-0644-5727]{Will Percival}
\affil{Department of Physics and Astronomy, University of Waterloo, 200 University Ave W, Waterloo, ON N2L 3G1, Canada}
\affil{Perimeter Institute for Theoretical Physics, 31 Caroline St. North, Waterloo, ON N2L 2Y5, Canada}
\affil{Waterloo Centre for Astrophysics, University of Waterloo, 200 University Ave W, Waterloo, ON N2L 3G1, Canada}

\author[0000-0001-7145-8674]{Francisco Prada}
\affil{Instituto de Astrof\'{i}sica de Andaluc\'{i}a (CSIC), Glorieta de la Astronom\'{i}a, s/n, E-18008 Granada, Spain}

\author[0000-0001-6979-0125]{Ignasi P\'erez-R\`afols}
\affil{Departament de F\'isica, EEBE, Universitat Polit\`ecnica de Catalunya, c/Eduard Maristany 10, 08930 Barcelona, Spain}

\author{Graziano Rossi}
\affil{Department of Physics and Astronomy, Sejong University, 209 Neungdong-ro, Gwangjin-gu, Seoul 05006, Republic of Korea}

\author[0000-0002-9646-8198]{Eusebio Sanchez}
\affil{CIEMAT, Avenida Complutense 40, E-28040 Madrid, Spain}

\author{David Schlegel}
\affil{Lawrence Berkeley National Laboratory, 1 Cyclotron Road, Berkeley, CA 94720, USA}

\author{Michael Schubnell}
\affil{Department of Physics, University of Michigan, 450 Church Street, Ann Arbor, MI 48109, USA}
\affil{University of Michigan, 500 S. State Street, Ann Arbor, MI 48109, USA}

\author[0000-0002-3461-0320]{Joseph H. Silber}
\affil{Lawrence Berkeley National Laboratory, 1 Cyclotron Road, Berkeley, CA 94720, USA}

\author{David Sprayberry}
\affil{NSF NOIRLab, 950 N. Cherry Ave., Tucson, AZ 85719, USA}

\author[0000-0003-1704-0781]{Gregory Tarl\'e}
\affil{University of Michigan, 500 S. State Street, Ann Arbor, MI 48109, USA}

\author{Benjamin A. Weaver}
\affil{NSF NOIRLab, 950 N. Cherry Ave., Tucson, AZ 85719, USA}

\author[0000-0001-5381-4372]{Rongpu Zhou}
\affil{Lawrence Berkeley National Laboratory, 1 Cyclotron Road, Berkeley, CA 94720, USA}

\author[0000-0002-6684-3997]{Hu Zou}
\affil{National Astronomical Observatories, Chinese Academy of Sciences, A20 Datun Road, Chaoyang District, Beijing, 100101, P. R. China}

\begin{abstract}

Using the Dark Energy Spectroscopic Instrument Milky Way Survey, we examine the six-dimensional space of the anticenter region of the Milky Way stellar disk (150$^\circ$ $<$ Galactic longitude $<$ 220$^\circ$) using 61,883 main-sequence turnoff stars. We focus on two well-known stellar overdensities in the anticenter: the Monoceros Ring (MRi) and Anticenter Stream (ACS). We find that the MRi overdensity has kinematic signatures consistent with a tidally induced spiral arm, a type of dynamic spiral arm created by an interaction with a satellite galaxy, most likely the Sagittarius dwarf spheroidal galaxy (Sgr). We use the kinematics of the MRi to calculate the two most recent passage times of Sgr are 0.25 $\pm$ 0.09 Gyrs and 1.10 $\pm$ 0.23 Gyrs from the present day. We validate that the ACS is kinematically decoupled from the MRi because they are moving in opposite radial and vertical directions. We find that the kinematics associated with the ACS extend beyond our defined overdensity. The features we see in the ACS region are likely part of a broader distribution of stars with the same kinematic signature as detected in other places, like the vertical wave in the outer disk and phase spiral.

\end{abstract}


\section{Introduction}
\label{sec:intro}

We have observational evidence that galaxies grow and evolve through major and minor mergers. This is an important test of our cosmological theory of hierarchical structure formation 
\citep{1980Dressler, 1992Barnes, 2006Baugh, 2010Benson, 2010aMoster, 2014Madau}. The internal structure of galaxies can be disrupted by these interactions and create dynamic components in disks, like spiral arms and bars. These components, in turn, influence the kinematics of the stars in the galaxy \citep{1972_Lynden_bell_kalnajs,2002Sellwood, 2018Erwin, 2025Zhang}.
The Milky Way (MW) is one example of a ``typical" late-type galaxy with spiral arms, a central bar, and evidence of hierarchical buildup in its stellar halo from merging satellite galaxies like the already accreted Gaia-Sausage-Enceladus (GSE), and the current accretion of Sagittarius dwarf spheroidal galaxy (Sgr) and the Large Magellanic Cloud (LMC) \citep{1994Ibata, 2007Besla, 2018Belokurov,2018Helmi, 2020Naidu, 2023Joshi, 2024Deason}. Stellar observations of the MW provide an accessible laboratory to explore galaxy interactions and accretion history as the kinematics of stars in a galaxy contain a record of their past, which allows us to understand how galaxies evolve over time.

The LMC and Sgr are currently the largest and most gravitationally significant galaxies interacting with the MW \citep{2018bLaporte_sgr_sim}. The LMC is on its first infall
\citep[][see also \citealt{2024Vasiliev} for an argument for the LMC being on its second infall]{2007Besla, 2013Kallivayalil, 2025arXiv25Lucchini}, 
Sgr has three proposed passages through the disk over the past eight billion years \citep{2010Law, 2015delaVega, 2017Dierickx, Laporte2019, 2020Ruiz_lara_nature}, and there is a tidal extension of stars belonging to Sgr forming a stellar stream \citep{1996Mateo, 2020Ramos}.
We can understand how the Galaxy is gravitationally influenced by satellites through observables such as the gravitational wake of the LMC \citep{ 2019Garavito_Camargo, 2021Conroy, 2025songtingli} and vertical perturbations in the disk \citep{2018bLaporte, 2018Antoja}.
Furthermore, elements such as the orbital history of Sgr, the progenitor's initial mass, and mass loss rate are intertwined such that it is difficult to precisely constrain one of these parameters without knowing the others. Using kinematics to deduce the pericenter passages of Sgr will advance future measurements of the mass of Sgr \citep{2018bLaporte}.

Spiral arms have many formation mechanisms, and spiral arms that manifest from gravitational instabilities from external sources, like satellite galaxies, are important to our understanding of disk structure evolution
\citep{2007Romero, BinneyTremaine2008, 2021Sellwood}.
Previous work has shown that satellite galaxy interactions can be directly linked to spiral arm formation \citep{1986Quinn,2010Dobbs, 2011Purcell, 2016gomez, Kawata2018, Khoperskov2022, 2022Carr, 2025Quinn}.
Due to our perspective in the Galaxy, we do not yet have a complete understanding of the MW's spiral structure, including the extent to which spiral arms appear in the outer disk 
\citep{2018bLaporte, Laporte2022, 2025Hunt}. Here we define the outer disk as galactocentric radii $>10$ kpc \citep{McMillan_2022}.
There may be many different processes at play to create the spiral structure of the Galaxy, therefore, it is important to identify which spiral arms exhibit the observational signatures of different types of spiral structure. Spiral arms induced by tidal forces from satellite galaxies are increasingly intriguing because they may also inform us about the accretion history of the Galaxy. They exhibit distinct kinematic features that can be used to investigate the initial pericenter passage time of the interaction \citep{1973Kalnajs,2022Antoja, 2024Stelea}, and these features are indicative of recent disk perturbations.

We currently observe the Galactic disk to be kinematically out of equilibrium, and this disequilibrium manifests as perturbations and asymmetries in the disk \citep[e.g.][]{Yanny_2013, McMillan_2022}. 
Observational evidence from the \textit{Gaia} space mission \citep{2016Gaia} of these disk perturbations is seen through the kinematics of the stars in multiple phase spaces, like in the ``phase space spiral" in vertical position and vertical velocity space \citep{Antoja2018, Bernet2022} and the diagonal ridges in Galactocentric radius and azimuthal velocity space \citep{Antoja2018, Ramos2018, 2023Antoja}.

Additionally, comparing observations to theoretical models of Galactic disk evolution may help us link the physics causing the disequilibrium to our observations. 
Recent studies from \cite{Fragkoudi2019}, \cite{Laporte2019}, and \cite{Bland_Hawthorn2021} show that N-body simulations can reproduce observed perturbations that we find in the MW disk with different internal or external mechanisms.
Two of the most dominant mechanisms driving Galactic disk perturbations are from internal and external sources. An example of internal sources is resonance orbits from the influence of the bar which mainly affect the inner disk and extends to the solar neighborhood \citep{1981Binney, 1984Pfenniger, Monari2019, Fragkoudi2019, Bernet2022}. An example of external sources is satellite galaxy interactions shaping the outer regions of the disk \citep{2018bLaporte,2020Eilers, 2022Antoja, Khoperskov2022}.

Along with the phase spiral and ridge patterns as evidence of disequilibrium, we have observed multiple overdense stellar regions in the MW outer disk.
The Monoceros Ring (MRi) was first discovered as a stellar overdensity by \cite{2002Newberg} using the Sloan Digital Sky Survey \citep[SDSS;][]{2000York}. The MRi region is located at heliocentric distances ($d$) $7 < d < 15$ kpc, and Galactic coordinates ($l$, $b$) ranging from $120^\circ < l < 260^\circ $, and $-30^\circ <b < 40^\circ $ \citep{2016Morganson, 2018Sheffield}.
The MRi was initially hypothesized to be accreted material of the remnants of a tidally disrupted dwarf galaxy \citep{2002Newberg, 2003Rocha_Pinto, 2004Martin, 2011MichelDansac, 2011Sollima}. More recent analyses demonstrate that MRi’s chemical composition, age distribution, and azimuthal velocity are constant with the MW disk \citep{2016gomez, 2020Laporte, Borbolato2024}. Instead of being made of accreted material, it is now well accepted that the MRi is made up of a population of MW stars.
The MRi may be MW disk stars forming this overdense region as a result of the gravitational influence of the infall of a satellite galaxy \citep{2018bLaporte, 2016gomez}.

The Anticenter Stream (ACS) is an overdense stellar region in the anticenter of the MW and was discovered by \cite{2006Grillmair}. Initially thought to be part of the MRi, we now know they are spatially separate in the sky \citep{Slater_2014, 2016Morganson}, and the two overdensities are moving in opposite radial and vertical directions \citep{2024Qiao}.
Since they lie in close proximity to each other, there have been debates on the plausibility of a physical connection between the MRi and ACS and their subsequent origins \citep{2006Grillmair, 2008Grillmair, 2008Kazantzidis, 2011Michel_Dansac,  Ramos_2021, 2020Laporte}.

Understanding the complete picture of these perturbations relies on the use of large stellar surveys like \textit{Gaia} \citep{2016Gaia}, Apache Point Observatory Galactic Evolution Experiment \citep[APOGEE;][]{2017Majewski},  Galactic Archaeology with HERMES  \citep[GALAH;][]{2015GALAH}, and Large Sky Area Multi-Object Fiber Spectroscopic Telescope \citep[LAMOST][]{2012LAMOST}, and now, the Dark Energy Spectroscopic Instrument (DESI) Milky Way Survey (MWS) \citep{2016aDESI, Cooper2023}.
These surveys provide insight into the motion and composition of stellar populations in the MW which one can use to uncover details of the kinematics of the disk.

In this paper, we examine the MRi and ACS regions using data from the first three years of DESI MWS. We show that the MRi overdensity possesses the same kinematic features as models of tidally induced spiral arms in the MW. Using the kinematics of the MRi overdensity, we can calculate the most recent pericenter passages of Sgr. The kinematics associated with the ACS region are different from the MRi region. We speculate that the ACS overdensity, instead of being a distinct stream of disk stars, could be part of a more broadly distributed population of stars that share the same kinematic signature which has been detected in other places, such as the phase spiral \citep{Antoja2018, 2023Antoja}. 
In Section \ref{sec:method}, we describe the DESI MWS, define the anticenter region where we focus our investigation, and outline the main-sequence turnoff star (MSTO) sample we use. In Section \ref{sec:results}, we show the motions of the MRi and ACS, indicating they are distinct kinematic groups. We compare our observations to previously studied N-body simulations of tidally induced spiral structures and calculate the most recent pericenter passage times of Sgr. We discuss our results in the context of previous work in Section \ref{sec:Discussion} and summarize our conclusions in Section \ref{sec:summary}.

\section{Data and Sample Selection}
\label{sec:method}
 
\subsection{DESI Milky Way Survey}

The Dark Energy Spectroscopic Instrument (DESI) is a multi-object, robotic, fiber-fed spectrograph operating on the Mayall 4-meter telescope at Kitt Peak National Observatory in Tucson, Arizona \citep{2022DESI}. It has 5,000 fibers and covers a wavelength range of 360 nm to 980 nm. It can simultaneously obtain spectra for nearly 5,000 objects over a $\sim$3$^\circ$ field of view \citep{2016DESI_sciencetargets, 2016bDESI, 2023Silber, 2023Guy_surveyops, 2023Schlafly_surveyops, 2024Miller, 2024Poppett}. The resolving power of the spectra varies between 2,000 at the blue end and 5,500 at the red end.
The DESI survey, started in May 2021, aims to obtain spectra from at least $\sim$ 63 million galaxies and quasi-stellar objects (QSOs) and $\sim$ 10 million stars in the ``MAIN-BRIGHT" program with a footprint of 17,000 square degrees in eight years, making it one of the largest spectroscopic surveys to date \citep{2016aDESI, Cooper2023}. 
The first data release \cite[DR1,][]{2025DESI_datarelease1} is now public and includes spectra for more than 18 million unique targets.
\cite{2025DESI_2024_VII} and \cite{2025Abdul_dr2results} show some examples of early cosmological results from DESI.

During dark sky conditions, DESI observations are designed to explore the physics governing the early universe via obtaining the most precise measurement of the expansion history of the universe to date using baryon acoustic oscillations \citep{2013Levi}.
The sky conditions to observe most of the high-redshift galactic targets must be dark and have good transparency. When observing efficiency across the DESI wavelength range is lower due to the phase of the moon, twilight, or poor seeing conditions, the survey switches to the MAIN-BRIGHT program, targeting stars in the Milky Way Survey \citep[MWS;][]{2020Allende, Cooper2023} and the Bright Galaxy Survey targets \citep[BGS;][]{2023Hahn}. 
The MWS is designed to investigate the local universe to better understand the accretion history of the MW and probe dark matter on small scales. This paper uses stellar spectra from the MWS MAIN-BRIGHT program. These targets were chosen based on Legacy Imaging Surveys \citep{2019Dey}, and the MWS target selection is described in \cite{Cooper2023}. The goal of the MWS is to measure radial velocities and chemical abundances for high galactic latitude stars that are fainter than the Radial Velocity Spectrometer (RVS) targets in \textit{Gaia}. For our analysis, this gives us a large new spectroscopic sample in the distant Galactic anticenter.

In this paper, we use MWS spectra from the first three years of DESI observations, which will be released as Data Release 2 (DR2) to the public in Spring 2027. \footnote{details in \url{https://data.desi.lbl.gov/doc/releases/}}. The data points corresponding to the figures from this paper are available at \url{http://zenodo.org/records/18236902}. The data reduction follows the DR1 stellar catalog described in \cite{2025Koposov}.  
The radial velocities are measured with the stellar parameter fitting code $\tt{RVSpecFit}$ \citep{2019Koposov}. The details of the pipeline are described in \cite{2024Koposov}. The parameters used in this paper and in DR1 were computed by an updated version of the pipeline that utilizes a neural network to interpolate between discrete stellar templates, which overcomes the issue of grid points in metallicity, surface gravity, and effective temperature \citep{2025Koposov}.
Accuracy and uncertainties of the radial velocities and metallicities are detailed in \cite{2025Koposov}.
The distances are calculated using the $\tt{rvsdistnn}$ method (Koposov et al., in prep) which is a neural network using stellar parameters like effective temperature, metallicity, color, and magnitude derived by $\tt{RVSpecFit}$. The formal uncertainties are estimated using Monte-Carlo sampling of the stellar parameter uncertainties and propagating them through the neural network. We compare the distances we use to published distances \cite[$\tt{SpecDis}$;][]{2025Li_specdis}. $\tt{SpecDis}$ is a value-added catalog published along with the DESI DR1 stellar catalog \footnote{\url{https://data.desi.lbl.gov/doc/releases/dr1/vac/mws/}}. Their distance measurements are derived from a neural network trained on the DESI MWS spectra along with \textit{Gaia} parallaxes and uncertainties on the parallax. For more details, we refer the reader to \cite{2025Li_specdis}.
The median ratio between the distances (distance$_{\tt{SpecDis}}$/ distance$_{\tt{rvsdistnn}}$) for MSTO stars in DESI DR2, is 0.98 with 1st and 3rd quartiles of 0.90 and 1.05, respectively, meaning the distances we use are in good agreement with the published data of \cite{2025Li_specdis}.
In the anticenter region for MSTO stars, our typical uncertainty in the Galactocentric radial coordinate, $R$, is 0.5 kpc at  $R = 12$ kpc and 2.0 kpc at $R = 18$ kpc. 

\subsection{Main-Sequence Turnoff Star Selection}

Because we want a high density of stars in the outer disk and avoid unnecessary foreground stars, we choose a sample of MSTO stars.
We start with the ``MAIN-BLUE" MWS sample, which is a subsample of the MAIN-BRIGHT program and selected in an $r$ magnitude range of 16 $<r<$ 19 mag and $g-r$ $<$ 0.7 \citep{Cooper2023}. Note that all magnitudes discussed are DESI Legacy Imaging Surveys magnitude bands and are extinction corrected using dust maps from \cite{SDF98}, and this selection does not apply any additional parallax or proper motion cut. 
We require the objects to be classified as ``STAR" by Redrock spectral classification \citep[Bailey et al., in prep,][]{2024Anand}, no warning flags for the radial velocity measurement 
($\tt{RVS\_WARN == 0}$), and a signal-to-noise ratio (SNR) in the $r$-band greater than 5. This leaves us with 2,989,951 stars.

\begin{figure}
    \centering
    \includegraphics[width=1\linewidth]{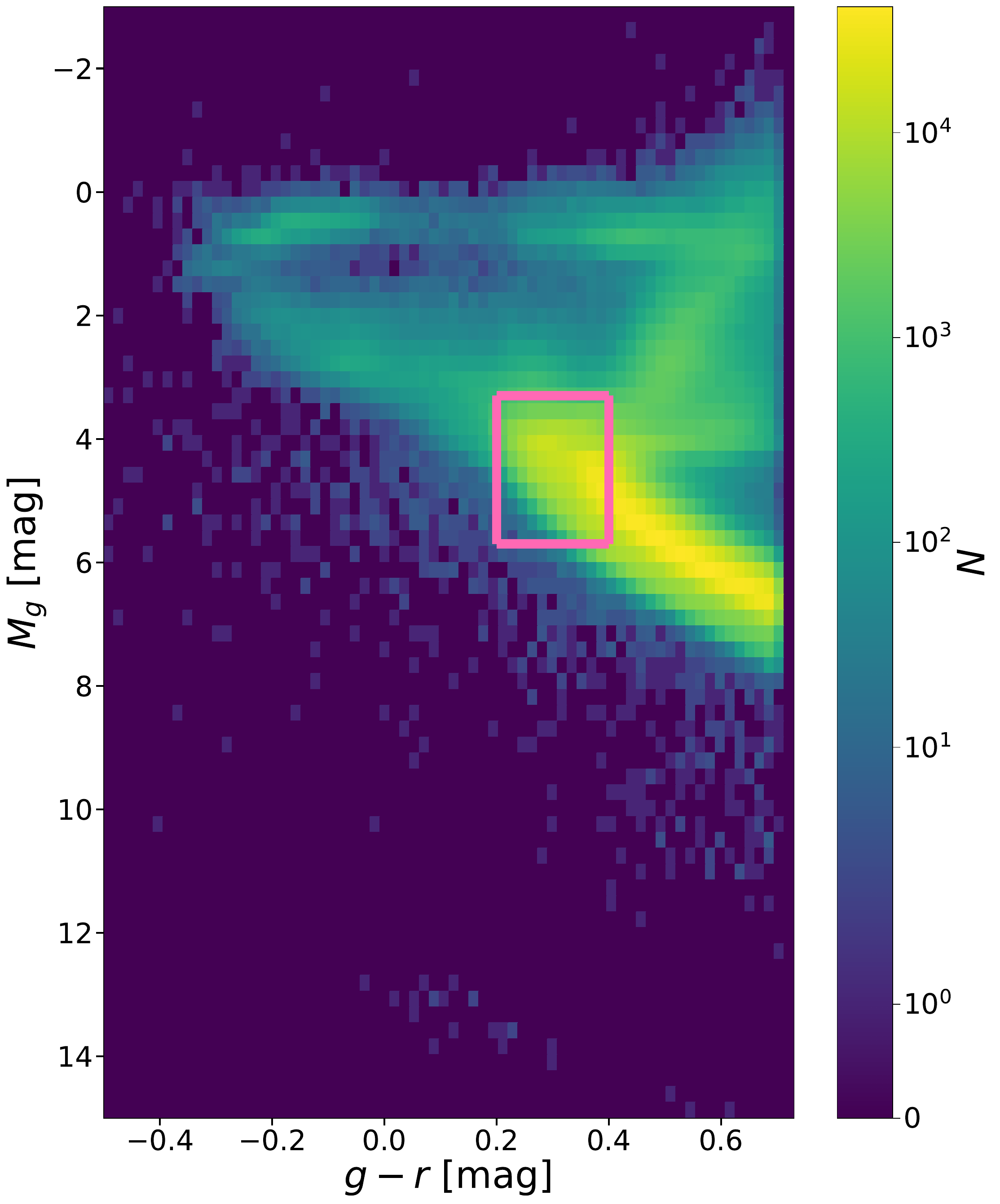}
    \caption{Two-dimensional histogram of the color--absolute magnitude diagram of the MWS MAIN-BLUE sample using 70 bins in each direction. The pink box shows our MSTO star selection, which is defined as $3.3<M_g<5.7$ and $0.2< g-r <0.4$.} 
    \label{fig:CMD}
\end{figure}

Figure \ref{fig:CMD} shows a color--absolute magnitude diagram of the stars in the MAIN-BLUE sample. We calculate the absolute magnitude corrected for extinction using the distance estimates described earlier. 
Our criteria for MSTO stars are $3.3<M_g<5.7$ and $0.2< g-r <0.4$. The pink box indicates our sample of MSTO stars.
Our MSTO selection gives us the number density advantage of using the stars on the main sequence while also avoiding unnecessary foreground stars. With our color selection of the region near the turnoff, we have a sample of stars at similar luminosity that reach the distances to the outer disk. 
We have 791,654 stars from these criteria.
We test increasing the color range of our MSTO sample to include redder stars, $0.2< g-r <0.5$.
We find that our results are consistent between the two color ranges, and this larger color range primarily increases the number of stars near the plane of the disk and are not close to the MRi or ACS region.
Therefore, we adopt the color range $0.2< g-r <0.4$ to avoid unnecessary foreground stars.

\subsection{Anticenter Sample}

\begin{figure}
    \centering
    \includegraphics[width=1\linewidth]{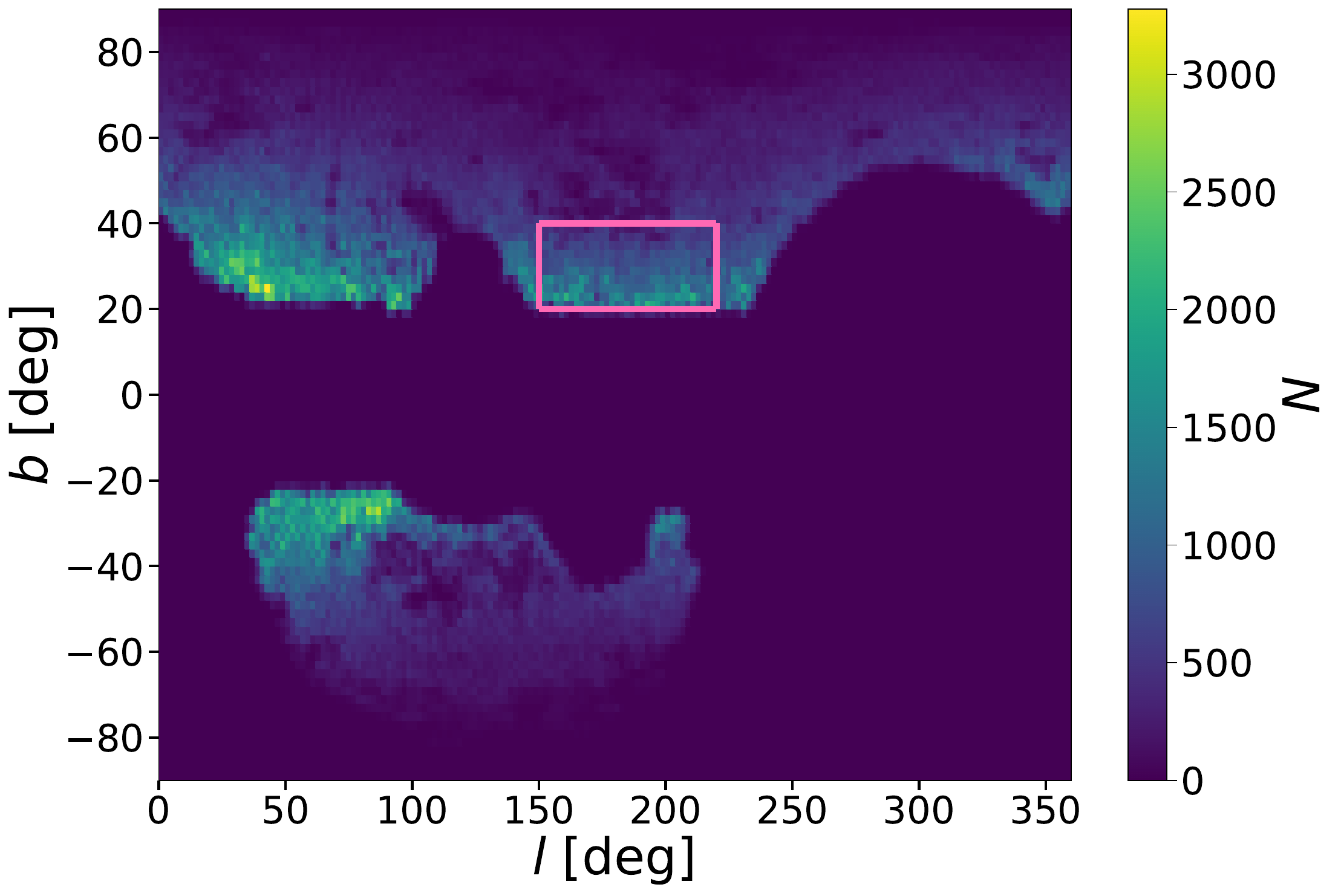}
    \caption{Density distribution of our MWS MAIN-BLUE sample in with 180 bins along $l$ and 90 bins along $b$. The pink box represents our anticenter sample coordinates between $150^\circ<l<220^\circ$ and $20^\circ<b<40^\circ$. We focus on the lowest latitude of the anticenter region, where we can observe the MRi in the DESI footprint and extending to the end of the ACS region that is above the DESI Galactic latitude limit. The gaps in data correspond to regions where the DESI MWS does not target due to high dust extinction and incompleteness in the survey. See Section \ref{completeness} for details.} 
    \label{fig:anticenterbox}
\end{figure}

Here, we define our sample in the anticenter region.  
In Galactic longitude ($l$), we define it to be between
$150^\circ<l<220^\circ$ and in Galactic latitude ($b$) we define it to be and $20^\circ<b<40^\circ$, as seen in Figure \ref{fig:anticenterbox}. The DESI MWS focuses on higher Galactic latitudes and avoids regions of high dust extinction \citep{Cooper2023}. Our anticenter sample extends to the lowest Galactic latitude in the DESI MWS footprint, includes as much of the MRi overdensity as is in the DESI footprint, and, at $l = 220^\circ$, reaches the end of the ACS region that is above the DESI Galactic latitude limit.

Using \textit{Gaia} DR3 proper motions along with distances, radial velocities, and sky coordinates from the DESI MWS, we calculate six-dimensional phase space coordinates for each star in our sample.
To create a sample of stars in the anticenter, we convert from ICRS to galactocentric cylindrical coordinates using Astropy \citep{astropy1, astropy2, 2022Astropy}.
Adopting a right-hand coordinate system, we take the distance from the Galactic center to the Sun to be 8.277 kpc \citep{2022GRAVITY} and its height above the Galactic plane is 0.0208 kpc \citep{2019Bennett}. We also take the Cartesian velocity vector of the Sun as (11.1, 248.5, 7.25) km s$^{-1}$ \citep{2010Schonrich, 2020Reid}.

\begin{figure}
    \centering
    \includegraphics[width=1.0\linewidth]{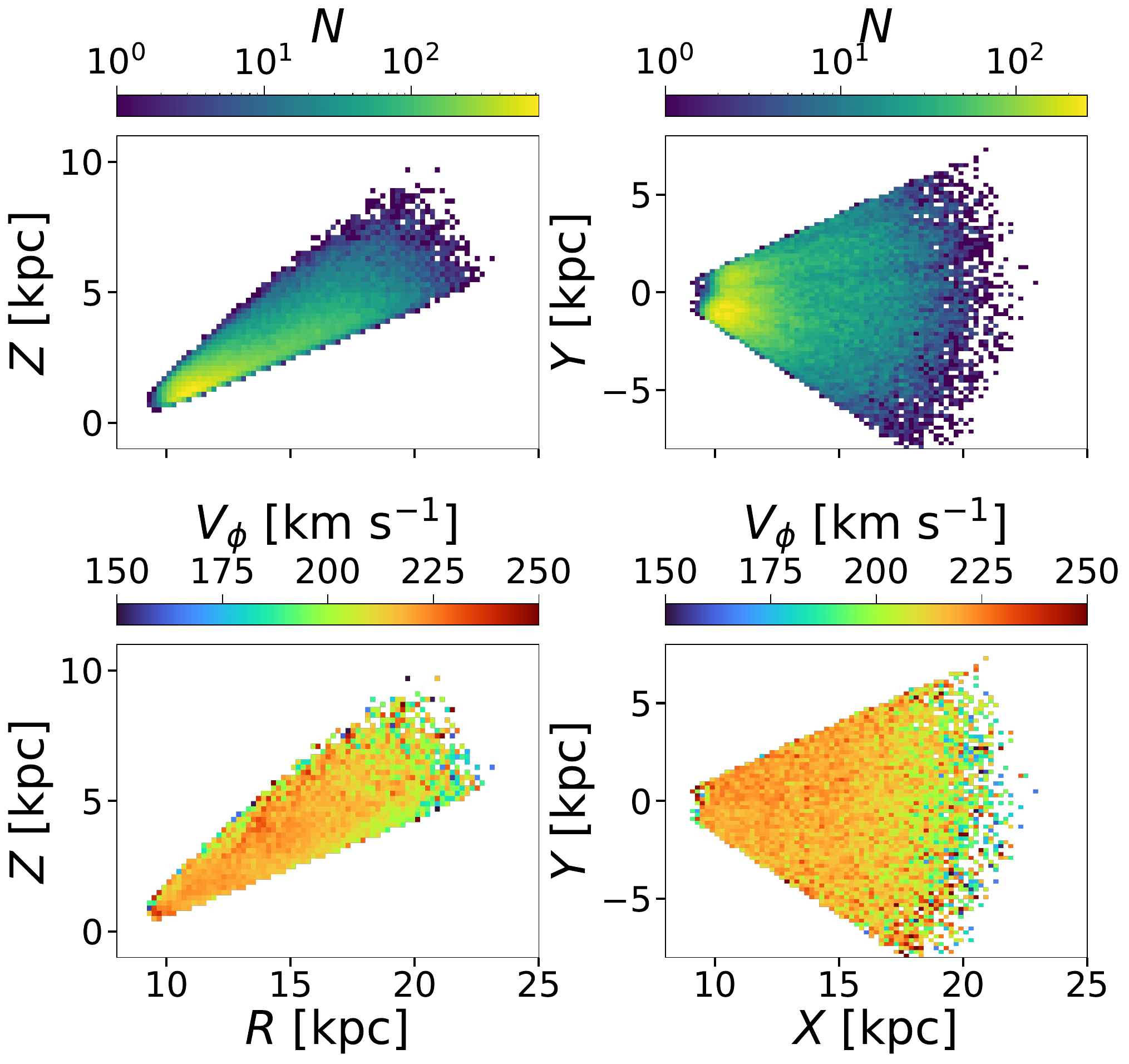}
    \caption{Left: Two-dimensional histogram of the $R$--$Z$ projection of the MSTO anticenter sample with five bins per kiloparsec in both directions. The top left panel shows the density distribution, and the bottom left panel is color-mapped to the median $V_\phi$ distribution. We see stars with comparable disk velocities extend to large $R$ and $Z$. Right: $X$--$Y$ projection using the same sample of stars as the left panels, with five bins per kiloparsec in both directions. The top right panel shows the density distribution of our sample, and the bottom right panel is color-mapped to the median $V_\phi$ distribution. We find that, after excluding stars with halo-like velocities, the median $V_\phi$ of this sample is 214 km s$^{-1}$, consistent with thin disk $V_\phi$ at the solar neighborhood.}
    \label{fig:R_Z_projection}
\end{figure}

We limit our analysis to galactocentric cylindrical radii between $8<R<25$ kpc. We also select stars with a vertical height ($Z$) between $0< Z <11$ kpc. The DESI  area covers much less of the anticenter region at $b<0^\circ$, so we do not compare kinematics above and below the Galactic plane in this paper. We make an azimuthal velocity ($V_\phi$) cut of 140 $<$ $V_\phi$ $<$ 277 km s$^{-1}$ in order to remove stars with halo-like velocities and a few stars with high velocities. As expected, the stars with halo-like velocities mostly occupied higher latitudes, $b\gtrsim35^\circ$.
Our final sample of MSTO stars in the anticenter is 61,883. We henceforth refer to this selection of stars as the MSTO anticenter sample.

Figure \ref{fig:R_Z_projection} illustrates stars in our MSTO anticenter sample in the $R$--$Z$ (left panels) and $X$--$Y$ (right panels) planes. The top row shows the density distribution, and the bottom row shows the median $V_\phi$ distribution.
We find that the median $V_\phi$ in our sample, after excluding stars with halo velocities, is 214 km s$^{-1}$, which is comparable to the $V_\phi$ at the solar neighborhood \citep{2012schonrish}. The median metallicity of our sample is [Fe/H]$=-0.74$ which is in agreement with disk metallicities \citep{2018Sheffield, 2020Laporte, Borbolato2024, 2024Qiao}. There is previous evidence of thin disk stars at these vertical heights; \cite{2023Han} and \cite{2024Uppal} show the scale height of the MW thin disk at $R\sim15$ kpc to be about 2.2 kpc. The large scale height in the outer disk of the MW is likely due to its recent accretion history \citep{1996Walker, 2018Sarkar, 2019Thomas}.

\subsection{Completeness}
\label{completeness}

\begin{figure}
    \centering
    \includegraphics[width=1\linewidth]{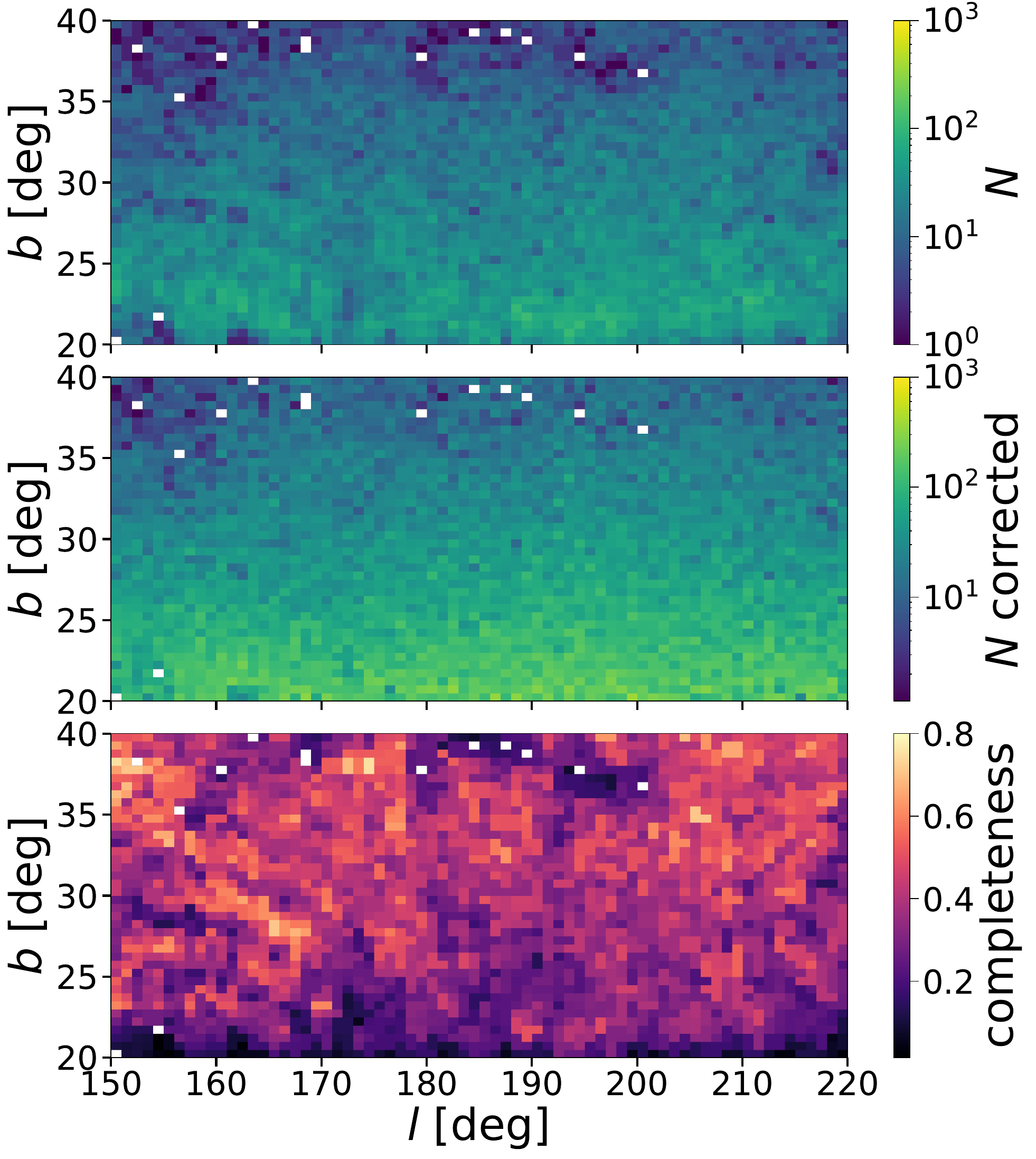}
    \caption{Top: Projection in $l$--$b$ space of the uncorrected density distribution of the MSTO sample with one bin per degree in $l$ and two bins per degree in $b$. 
    Middle: Same as above, except showing the density distribution corrected for incompleteness. 
    Bottom: Same as above, mapped to the completeness ratio. We see that the middle panel is more homogeneous than the top panel due to the completeness correction.}
    \label{fig:corrected_completeness}
\end{figure}

Because we are using data from the first three years of DESI operations out of the eight planned years, there are still targets that have yet to be observed, and density variations seen in the sample may be due to this variable completeness and not physical overdensities (see top panel of Figure \ref{fig:corrected_completeness}).
We correct for this incompleteness in order to measure consistent relative densities in the anticenter region. 

We correct for observational incompleteness by comparing the density of stars in the DESI target catalog to our spectroscopic sample. The full DESI target catalog was fixed at the start of the survey. It was drawn from the Legacy Survey \citep{2019Dey} using the selection criteria described in \cite{Cooper2023}.  
We divide our anticenter region into bins of one degree in Right Ascension (RA) and Declination (Dec). 
For each bin, we compute the ratio of the number of observed MAIN-BLUE stars divided by the total number of MAIN-BLUE stars in the target catalog. We can assume the the completeness of our MSTO sample is similar to the completeness of the MAIN-BLUE targets.

This completeness ratio is assigned to all the stars located in that bin (bottom panel of Figure \ref{fig:corrected_completeness}).
To implement this correction, we divide the number of stars by the completeness ratio assigned to each star, which returns a completeness-weighted number of stars (middle panel of Figure \ref{fig:corrected_completeness}). We verify that we recover the on-sky target numbers after the completeness correction. Because dust extinction reduces the original number of targeted stars in the MAIN-BLUE sample, the number of targets is not the same as the total number of stars on-sky. Incompleteness corrected number density will be labeled as ``N corrected'' in subsequent figures.
The upper region of the top and middle panels of Figure \ref{fig:corrected_completeness} shows a few bins with very low density due to dust extinction which do not affect our results.  

\section{Results}
\label{sec:results}

\subsection{Kinematics of the MSTO Anticenter Sample}
\label{sec:ac_results}

\begin{figure*}
    \centering
    \includegraphics[width=1.0\linewidth]{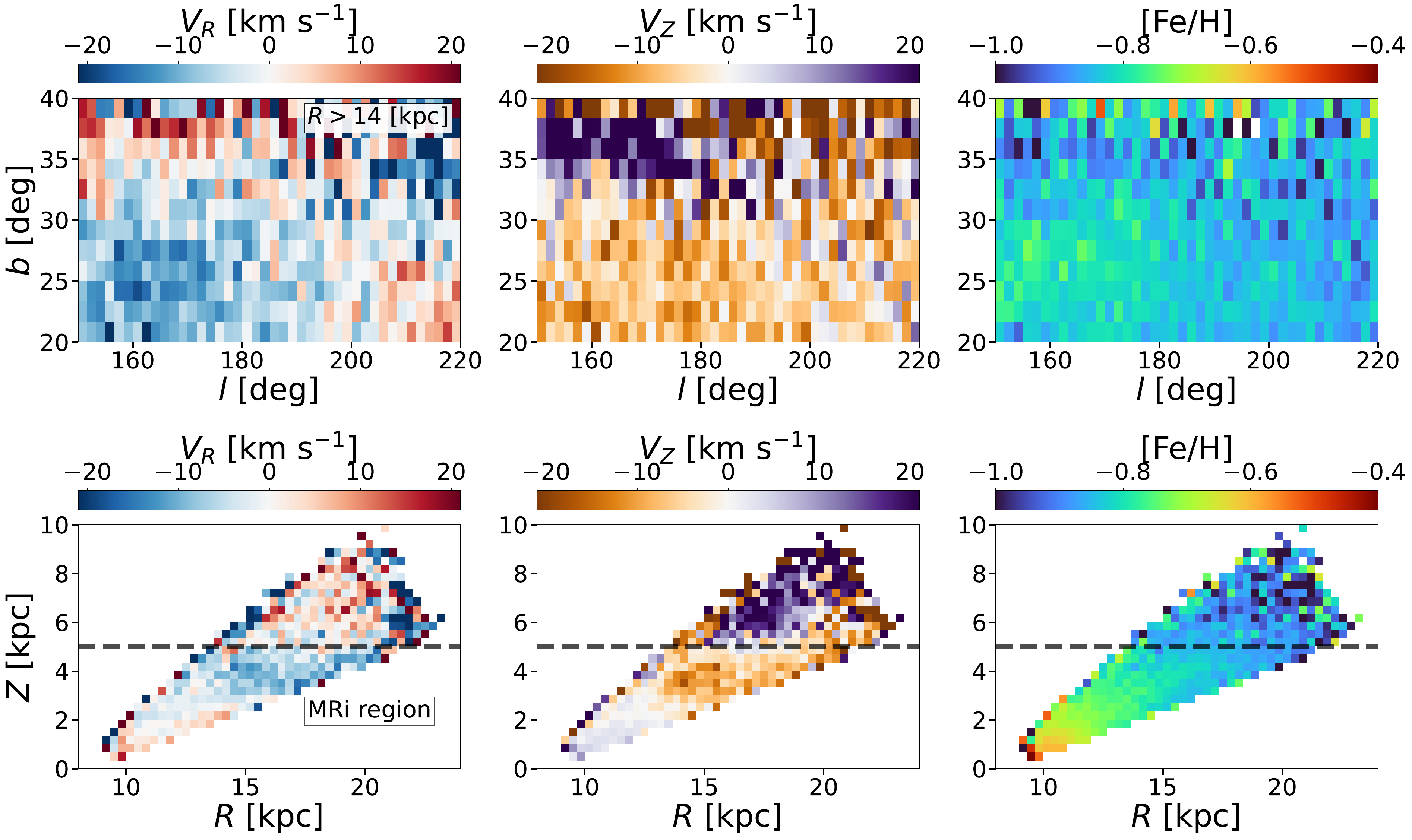}
    \caption{Top: $l$--$b$ projection of the MSTO anticenter sample with $R>14$ kpc where color is mapped to the median value of $V_R$ (left), $V_Z$ (middle), and [Fe/H] (right) with one bin per degree along both axes. Bottom: $R$--$Z$ of the MSTO anticenter sample with the same colormapping as above. We see distinct kinematic groups in $V_R$ and $V_Z$ where $-V_R$ and $-V_Z$ regions (illustrated in blue/yellow, respectively) are located in the region associated with the MRi overdensity. The $+V_R$ and $+V_Z$ regions (illustrated in red/purple, respectively) correspond to the ACS region. The horizontal black dashed line shows where $Z=5$ kpc and indicates our defined separation between the MRi region and other higher $Z$ stars.}
    \label{fig:lb_RZ}
\end{figure*}

The top row of Figure \ref{fig:lb_RZ} illustrates the MSTO anticenter sample ($150^\circ<l<220^\circ$ and $20^\circ<b<40^\circ$) projected in Galactic longitude and latitude with one bin per degree along both axes. To focus on the region where the MRi and ACS are found, we limit the sample to the outer disk region where $R > 14$ kpc. The bottom row of Figure \ref{fig:lb_RZ} shows the MSTO anticenter sample in the $R$--$Z$ plane with three bins per kiloparsec. The left column colors map to the median radial velocity ($V_R$) in each bin, the middle column maps to the median vertical velocity ($V_Z$), and median [Fe/H] is mapped in the right panel. We see two distinct kinematic groups in these panels.
First, in the top row, we see a region of $-V_R$ (blue) and $-V_Z$ (yellow) between 150$^\circ$ $<l<$ 200$^\circ$ and 20$^\circ$ $< b <$ 33$^\circ$, and it appears that the $-V_Z$ region extends in longitude to about 220$^\circ$. Secondly, the regions of $+V_R$ (red) and $+V_Z$ (purple) span approximately the same range in coordinates between 150$^\circ$ $< l <$ 190$^\circ$ and 33$^\circ$ $<b<$ 38$^\circ$. 

The $-V_R$ and $-V_Z$ regions in the top panels are the same as the regions with the $-V_R$ and $-V_Z$ motions in the region from $14< R <20$ kpc and $3\lesssim Z <5$ kpc in the bottom panels. The $+V_R$ and $+V_Z$ regions in the top panels are the same as the regions with the $+V_R$ and $+V_Z$ motions in the region from $15< R <21$ kpc and $Z >5$ kpc in the bottom panels. The regions of $-V_R$ and $-V_Z$ motions are located in the part of the Galaxy associated with the MRi ($Z<$ 5 kpc). 
The $+V_R$ and $+V_Z$ motions are located in the region loosely associated with the ACS. 
We refer the reader to Figure \ref{fig:l_b_panstarrs} for an illustration of the photometric density in this region.

Previous studies have found that the MRi is located between $14<R<20$ kpc \citep{2023Xu, Borbolato2024, 2024Qiao}. 
We find that the MRi region is centered at $R\sim16$ kpc.
This is on the closer end of the range, but we note that our Galactic latitude limit cuts off more distant parts of the MRi at low latitude, as seen in the bottom row of Figure \ref{fig:lb_RZ}.
Our kinematic maps in Figure \ref{fig:lb_RZ} are consistent with the work of \citet{2020Laporte} and \citet{2024Qiao}, arguing that the MRi and ACS are both overdensities in the disk but are separate and distinct because they have opposite radial and vertical motions and are kinematically decoupled.

The right column of Figure \ref{fig:lb_RZ} shows that the metallicities of the MSTO anticenter region are similar to metal-poor thin disk stars \citep{2018Sheffield, 2020Laporte, Borbolato2024, 2024Qiao}. 
The black horizontal dashed line in the panels of the bottom row is drawn at $Z =$ 5 kpc. We use this to separate the MRi region in the lower half from higher Z stars that may be associated with the ACS region in the upper half.
For $Z<$ 5 kpc, we measure a median [Fe/H] of $-$0.73.
In Section \ref{kine_ACS}, we apply a specific ACS selection and find the median metallicity to be $-$0.87.
The ACS region is more metal-poor than the MRi region which is consistent with \cite{2020Laporte} and \cite{2024Qiao}.

At the distances of the MRi and ACS, the average line-of-sight velocity error is $\sim$ 3 km $s^{-1}$. We estimate the transverse velocity error is $\sim$ 6 km $s^{-1}$ per coordinate from the contribution of the proper motion and distance uncertainties.
Because the bottom left and middle panels of Figure \ref{fig:lb_RZ} show the median velocities per bin and there are roughly 200 stars per bin at the MRi and ACS region, the uncertainties on the bin values are smaller than the features seen in the color maps.

\subsection{The Monoceros Ring Overdensity}
\label{sec:MRi_overdensity}

\begin{figure*}
    \centering
    \includegraphics[width=1.0\linewidth]{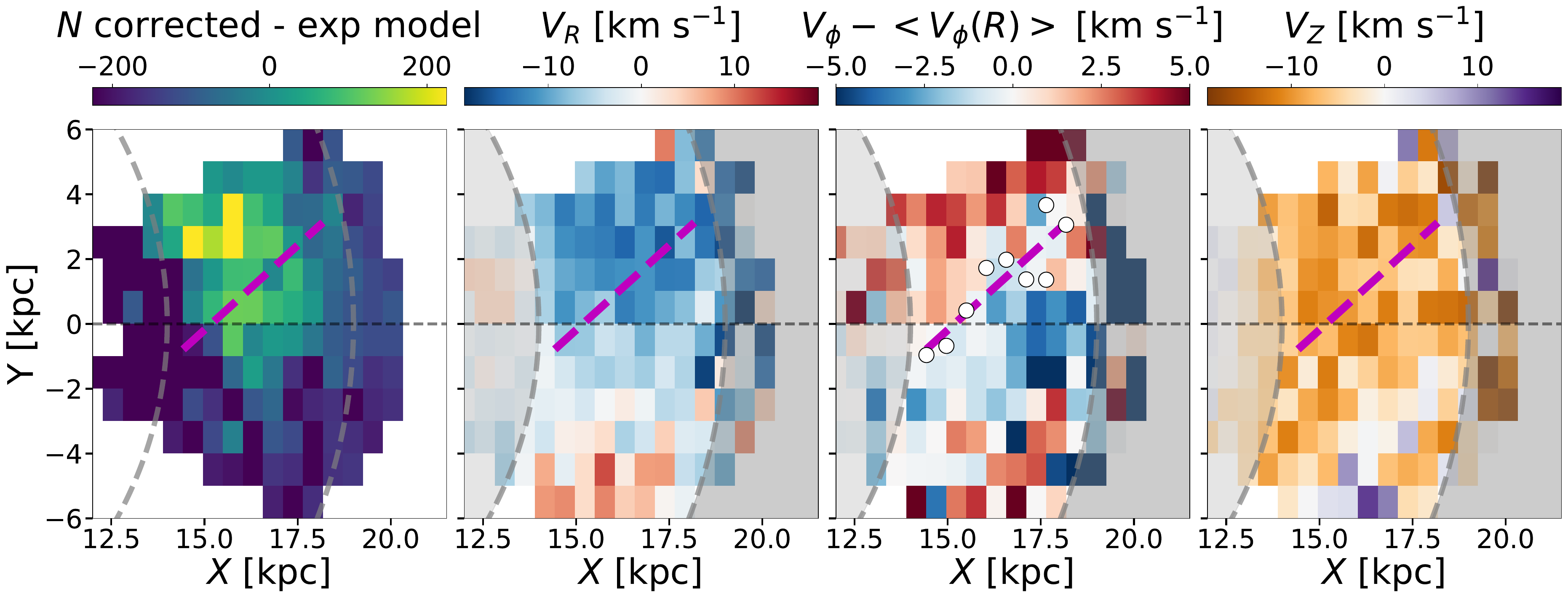}
    \caption{Left: $X$--$Y$ plane of the relative density distribution of MSTO anticenter sample with $0< Z < 5$ kpc computed by subtracting a decreasing exponential function modeling the stellar density of the disk with two bins per kiloparsec in $X$ and 1 kpc per bin in $Y$. Second panel from left: $X$--$Y$ distribution of stars in the MSTO anticenter sample with color mapped to the median $V_R$ in each bin. The large $-V_R$ (blue) region correlates with the position of the MRi overdensity. Third panel from left: the same except mapped to the median $V_{\phi} - <V_\phi$($R$)$>$. Right: the same except mapped to $V_Z$ with the $-V_Z$ (yellow) region corresponding to the MRi.
    Only the left-most panel includes a disk model subtraction. The black dashed horizontal line indicates $Y= 0$ kpc ($l=180^\circ$), while the dashed curved lines indicate $R=$ 14 and 18 kpc. The magenta dashed line represents the best-fit line to the inflection points of $V_\phi - <V_\phi$($R$)$>$ from 14 $<R<$ 19 kpc (white dots in the third panel). Our observations indicate that the localized $V_\phi$ inflection line, indicating corotating motion, is coincident with the $-V_R$ region and the overdensity associated with the MRi. This means the MRi overdensity is consistent with the kinematic signatures of a tidally induced spiral arm \citep{2022Antoja}. See text for more details.}
    \label{fig:XY_vr_vphi_vz}
\end{figure*}

\begin{figure}
    \centering
    \includegraphics[width=1\linewidth]{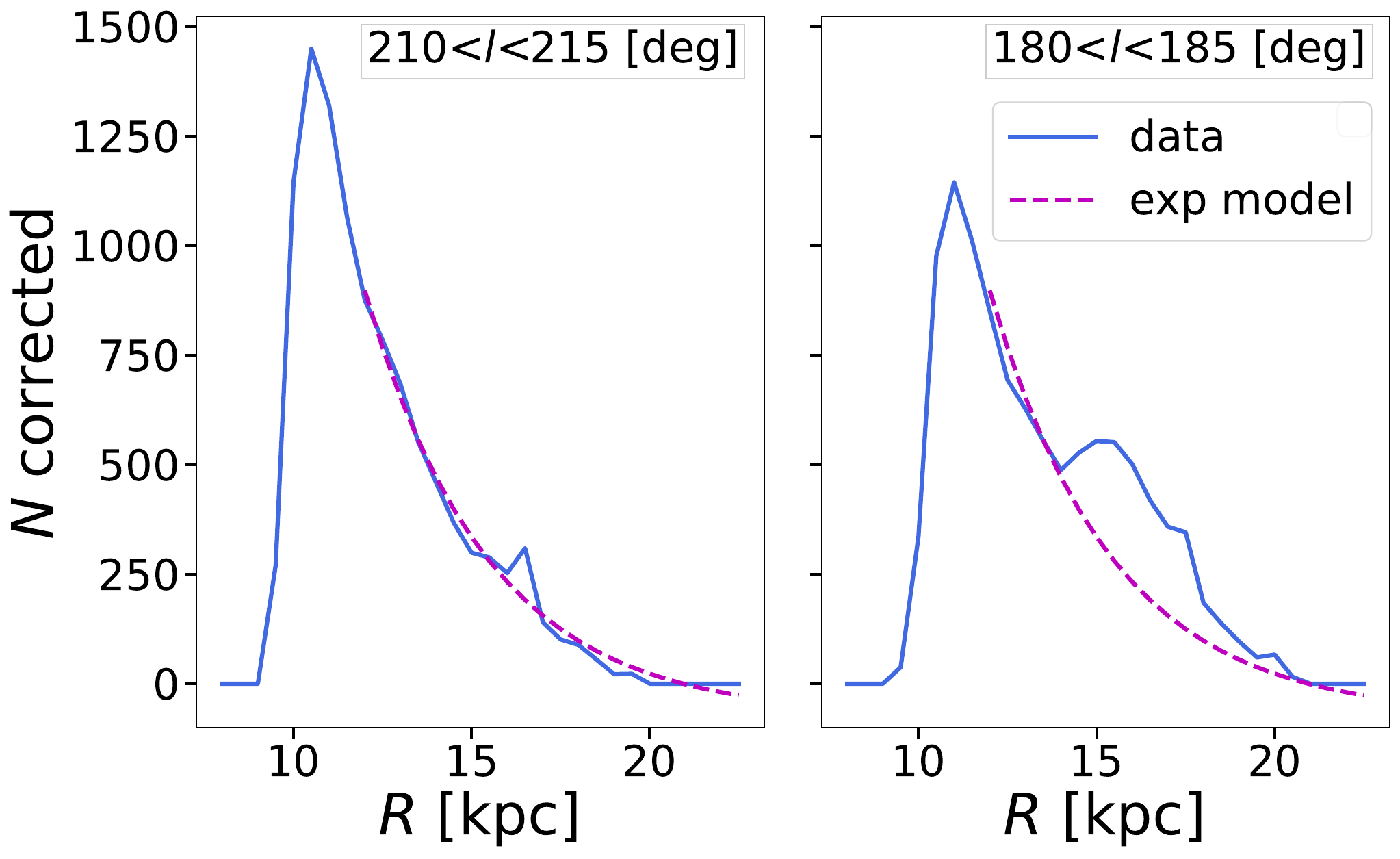}
    \caption{Completeness-corrected star count versus $R$ for the MSTO anticenter sample in a range between $210^\circ<l<215^\circ$ (left) and  $180^\circ<l<185^\circ$ (right) and in blue. The magenta dashed line is a decreasing exponential function fit to the data between $210^\circ<l<215^\circ$ overlaid in both panels. We see the exponential function matches well to the left panel, the region where we do not expect to see the MRi overdensity to be located, while the right panel shows an overdense region at $R\sim16$ kpc.} 
    \label{fig:modeling_panels}
\end{figure}

To define the specific regions we are discussing, we use the term “MRi region” in a broad sense to describe the stars within a $0<Z<5$ kpc cut in the MSTO anticenter sample (as shown by the dashed line in Figure \ref{fig:lb_RZ}). This region includes both the MRi overdensity as well as disk stars not associated with the overdensity. This allows us to compare the kinematics between the MRi overdensity stars and other disk stars not associated with the overdensity. We use the term ``MRi overdensity” to refer to the stars in the overdense region we find in the left panel in Figure \ref{fig:XY_vr_vphi_vz} between $14<R<18$ kpc. 
Further in this section, we compare our analysis to previous analyses of the MRi overdensity using MSTO anticenter stars selected via isochrones which were historically used to select member stars of the MRi overdensity \citep{2018deboer, 2016Morganson}.

Figure \ref{fig:XY_vr_vphi_vz} shows our MRi region selection in the $X$--$Y$ plane with two bins per kiloparsec in $X$ and one bin per kiloparsec in $Y$. We remove bins that contain five or fewer stars, most of which are located at $R>$ 20 kpc. The black dashed horizontal line indicates where $Y = 0$ kpc. The dashed curved lines indicate $R =14$ and 18 kpc, which bound the main part of the MRi overdensity.

The left panel of Figure \ref{fig:XY_vr_vphi_vz} shows the residual number density after subtracting a smoothly decreasing exponential model of the stellar disk's density. 
We fit a simple decreasing exponential model to the 0 $<Z<$ 5 kpc cut MSTO anticenter sample in a range of $210^\circ< l<215^\circ$. Our model's projected surface density is $\propto \exp\frac{-R}{h_R}$ where $h_R$ is a best-fit scale length of 3.46 kpc. This is illustrated in the left panel of Figure \ref{fig:modeling_panels}. In this region, the density is reasonably well described by a decreasing exponential function, and there is almost no overdensity present. The right panel of Figure \ref{fig:modeling_panels} shows an example of our exponential model compared to this sample in a range between $180^\circ<l<185^\circ$ ($l=180^\circ$ is equivalent to $Y=0$ kpc), and we see an overdensity at $R\sim16$ kpc.

We apply the same exponential model to the whole sample in bins of five degrees in $l$ from $150^\circ < l < 220^\circ$, normalized so that the model matches the data at $R=12$ kpc. Then, we calculate the residual between the data and the model as seen in the left panel of Figure \ref{fig:XY_vr_vphi_vz}. 
We see an overdensity in the $180^\circ<l<185^\circ$ region where we expect the MRi to be located, whereas the $210^\circ <l< 215^\circ$ region shows the data located away from the main part of the MRi overdensity. We note that our density analysis is only supplementary to our kinematic analysis, and therefore we do not attempt to make precise relative density measurements in the MRi region. Figure \ref{fig:modeling_panels} and the left panel of Figure \ref{fig:XY_vr_vphi_vz} demonstrate that we can identify the MRi overdensity in our data, highlighting the region where our kinematic analysis will focus. 

The MRi overdensity in the left panel of Figure \ref{fig:XY_vr_vphi_vz} is centered at $R \sim 16$ kpc. It is offset toward positive $Y$ values and appears to extend to the edge of our sample in Galactic longitude. This non-centered overdensity also corresponds to the MRi region in the top row of Figure \ref{fig:lb_RZ} where the $-V_R$ and $-V_Z$ region runs up against our $l$ cut at 150$^\circ$ while trailing off at $l=195$$^\circ$.
Photometric maps of the MRi find it extends in Galactic longitude beyond our sample to $l\sim$ 120$^\circ$ \citep{2016Morganson}.
We associate this overdensity with the MRi because the radial and vertical kinematics within the overdensity shown in Figure \ref{fig:XY_vr_vphi_vz} also appear in the same location in the Galaxy as found in previous work \citep{2020Laporte, 2023Xu,2024Qiao}.
The kinematic panels of Figure \ref{fig:XY_vr_vphi_vz} (the second, third, and fourth panels from the left) do not include a disk model subtraction. We do not subtract the stellar density model of the disk in order to show the kinematics of the MRi in contrast to the surrounding disk region.

Previous work investigating the MRi used isochrones to select the population of stars from photometric data \citep{2018deboer, 2016Morganson}. We verify that when we use a Dartmouth Stellar Evolution isochrone \citep{2008Dotter} with an age of 8 Gyrs,
and [Fe/H] of $-0.95$ \footnote{assuming the Sun's metal mass fraction is 0.019 \citep{2024Buldgen}}, the same isochrone used in \cite{2018deboer}, at a distance of 10 kpc to select a MRi sample, we find an overdensity in the same location in the disk with the same kinematics as the MRi overdensity in Figure \ref{fig:XY_vr_vphi_vz}.
Because \cite{2018deboer} uses a photometric sample with a fainter magnitude limit, their sample includes more distant stars at lower $b$ compared to our sample's limit at $b > 20^\circ$ (Figure \ref{fig:lb_RZ}), whereas we only see the tip of the MRi main-sequence turnoff in our spectroscopic MSTO sample.
We also verified that when we use an isochrone with similar properties to \cite{2018deboer}
(age of 9 Gyrs, [Fe/H] of $-0.84$,
and distance of 8 kpc),
which is a better match to the part of the MRi main sequence above the magnitude limit of our spectroscopic sample, we still select stars with similar properties to the MRi region in Figure \ref{fig:XY_vr_vphi_vz}. 
However, for our kinematic maps, we include all stars with $Z<$ 5 kpc to compare the MRi to the surrounding stars.

\subsection{Kinematics of the Monoceros Ring Region}
\label{sec:MRi}

\subsubsection{Radial and Azimuthal Velocities of the Monoceros Ring Region}
The second panel from the left of Figure \ref{fig:XY_vr_vphi_vz} is color-coded by the median $V_R$ in each bin, and shows the strong $-V_R$ region ($\sim$ $-$10 km s$^{-1}$) in the same location as the MRi overdensity in the left-most panel.
Observations with APOGEE in the MRi region from \cite{2020Eilers} and \cite{2024Qiao} also show a strong negative radial velocity signal at $R\sim 16$ kpc, which is qualitatively consistent with Figure \ref{fig:XY_vr_vphi_vz}.

The color in the third panel from the left maps to the difference between $V_\phi$ and the average $V_\phi$ at a given radius ($V_\phi - <V_\phi$($R$)$>$). This is calculated by measuring the average $V_\phi$ in 0.5 kpc annuli in $R$ and subtracting that from the $V_\phi$ values for each star. The median value of the difference is plotted for each bin.
In the region associated with the MRi, we see the velocities transition from faster than average (red) to slower than average (blue). We fit a cubic function to the MRi region in each annulus in $R$ 
to find where $V_\phi - <V_\phi$($R$)$>$ is closest to zero, which is the location where the stars are moving at the average $V_\phi$ at each radius. We plot that location with white circles in the $V_\phi - <V_\phi$($R$)$>$ panel.
The magenta dashed line is the best-fit line describing where $V_\phi - <V_\phi$($R$)$>$ is closest to zero in each radial annulus, the inflection point of $V_\phi - <V_\phi$($R$)$>$ from our cubic fit. The equation for the magenta line is $Y = 1.04 \cdot X -15.84$ and is overplotted in all four panels. We note that the inflection line is also coincident with the $-V_R$ region and is within the MRi overdensity. 
The average $V_\phi - <V_\phi$($R$)$>$ values at $Y\sim-4$ kpc and $X\sim17$ kpc have large positive and negative fluctuations and do not show the same coherent trend as the values at $Y > 0$ kpc that are associated with the MRi overdensity. 

\subsubsection{Monoceros Ring Connection to Tidally Induced Spiral Arms}
The association of the $-V_R$ region and inflection line in $V_\phi - <V_\phi$($R$)$>$ is a characteristic trait of tidally induced spiral arms \citep[Figure 2;][see also \citealt{1973Kalnajs}]{2022Antoja}. 
Tidal interactions between the MW and Sgr-like galaxies produce spiral structure in disks \citep{2011Purcell, 2016gomez, Kawata2018, Khoperskov2022, 2022Carr}. The signature of these tidally-induced spiral arms is that the minimum (most negative) $V_R$ occurs near the peak of the arm overdensity \citep{2022Antoja, 2024Stelea}. At the same location, there is an inflection point in $V_\phi - <V_\phi$($R$)$>$. This is because the arm overdensity is corotating, so the center of the arm is moving at the local $V_\phi$ \citep{2022Antoja,2024Stelea,1973Kalnajs}.

\cite{2022Antoja} models the interactions between MW-like and Sgr-like galaxies in the plane of the disk as an impulsive, weak interaction that gives velocity kicks in $V_R$ and $V_\phi$ that are slow compared to the rotation speed. This shows that the overdensity of a tidally induced spiral arm corresponds to a region of strong $-V_R$ and local $V_\phi$ inflection lines.
\cite{2022Antoja} applies the velocity kicks in $V_R$ and $V_\phi$ and chooses arbitrary values of $\Delta V=10$ kpc and $D=20$ kpc, where $D$ is a scale parameter of the radius of the disk. 

Conceptually, in the plane of the disk, a satellite passing through or near the host's disk will give a velocity kick to the stars, causing their orbits to rotate with respect to each other. This leads to orbital crowding and the formation of a spiral arm pattern in the regions where the orbits are moving towards pericenter \citep{2022Antoja}. This demonstrates why we see the strong kinematic signatures in $V_R$ and $V_\phi - <V_\phi$($R$)$>$ aligned with the modeled spiral arms in \cite{2022Antoja} and as seen in the overdensity in Figure \ref{fig:XY_vr_vphi_vz}. Our maps of the kinematic signatures of the MRi region seen in Figure \ref{fig:XY_vr_vphi_vz} are consistent with the kinematics of the tidally induced spiral structure seen in \cite{2022Antoja} and \cite{2024Stelea}.
In \cite{2024Stelea}, they find that, at $R\gtrsim$ 15 kpc, Sgr and the LMC are likely to be the dominant influence on outer disk structure, whereas secular processes can be more significant in the inner disk.
If the spiral overdensity pattern rotates at different rates as a function of radius, which we expect for the nearly flat rotation curve in the MW, the spiral arms wind up with time. This makes the arms an observable kinematic signal for the initial pericenter passage time of the satellite interaction. We investigate the orbital history of Sgr in Section \ref{sec:timing}.

\subsubsection{Vertical Velocities in the Monoceros Ring Region}

The right panel of Figure \ref{fig:XY_vr_vphi_vz} is color coded to the median vertical velocities of this region, and we see strong $-V_Z$ velocities in a similar location to the strong $-V_R$. The strong $-V_Z$ region appears to trail towards negative $Y$ values at a slightly closer radius. Observations with APOGEE in the MRi region from \cite{2024Qiao} also show a strong negative vertical velocity signature at $R\sim 16$ kpc, which is qualitatively consistent with Figure \ref{fig:XY_vr_vphi_vz}.
Previous studies, such as \cite{2012Widrow} and \cite{2013Williams} have also found vertical oscillations in the MW disk and speculate that they may have been caused by gravitational interactions between the Galaxy and Sgr. \cite{2015Xu} and \cite{2018Sheffield} find the MRi region may be related to vertical oscillations in the outer disk. 
\cite{2022Antoja} does not analyze $V_Z$ motion in their tidally induced spiral arm model. 
We instead compare the vertical velocities of the MRi region with the N-body simulations of tidal interactions between the MW, Sgr, and LMC in \cite{2024Stelea}. 
In the simulation by \cite{2024Stelea}, we look at velocity features that indicate tidal disruption caused by Sgr. The vertical motions at $R\sim 16$ kpc of our results are consistent in magnitude and direction with Figure 8 in \cite{2024Stelea}. However, they do not look at stars above $Z>1$ kpc. 

\begin{figure*}
    \centering
    \includegraphics[width=1.0\linewidth]{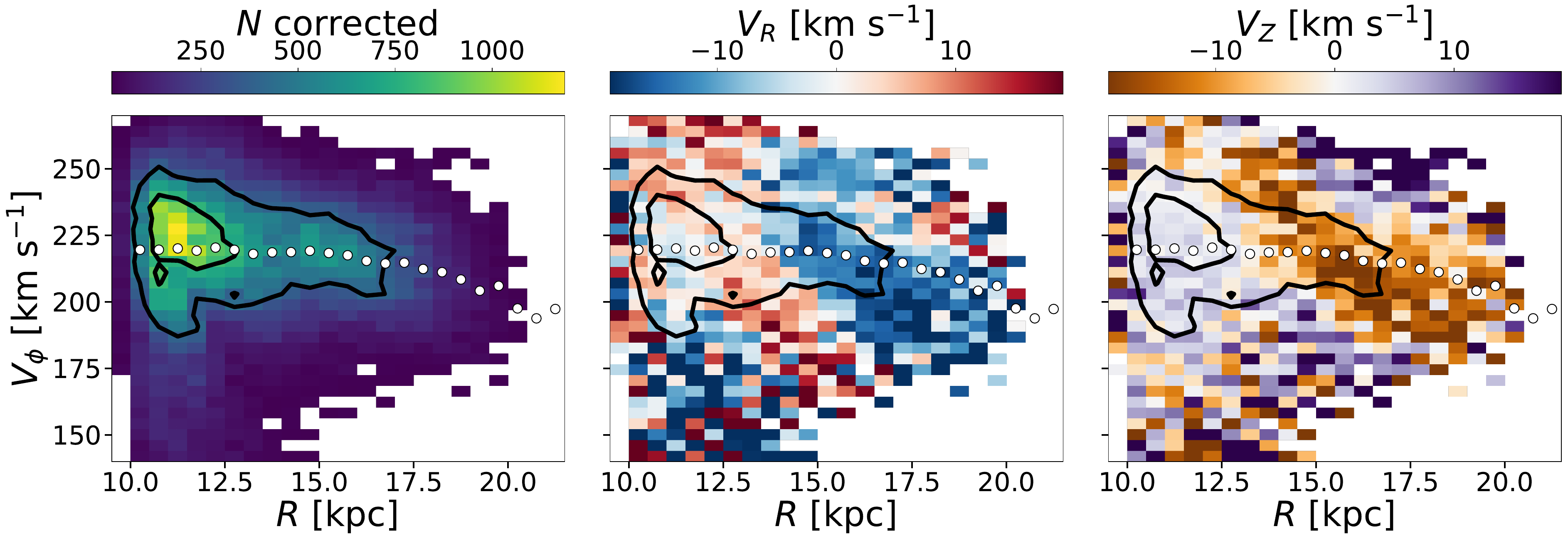}
    \caption{Left: $R$--$V_\phi$ phase space of the density distribution of stars in the MRi selection from the MSTO anticenter sample above $Y>0$ kpc with 2 bins per kiloparsec in $R$ and 0.25 bins per kilometers per second in $V_\phi$. Middle: $R$--$V_\phi$ with the same binning as the left panel color-mapped to median $V_R$ per bin. Right: Same as left panel but with color mapped to median $V_Z$. We remove bins with less than 20 stars, and density contours are illustrated in black. The white circles show the median $V_\phi$ at each $R$. We find that the median $V_\phi$ per $R$, indicating the location of corotating stars, corresponds to the $-V_R$ and $-V_Z$ motions in the densest region of the figure.
    }
    \label{fig:R_V_phi_VR_VZ}
\end{figure*}

In summary, we find the kinematics in the MRi overdensity, $-V_R$ and ($V_\phi - <V_\phi$($R$)$>) \sim 0$, are distinct from the rest of the kinematics in this region. When comparing these features to recent models from \cite{2022Antoja}, these observed kinematics are consistent with the peak overdensities of models of tidally induced spiral arms caused by an impulsive flyby of Sgr. Our observed kinematics in $V_R$ and $V_Z$ are also consistent with the simulations produced by \cite{2024Stelea}, where Sgr and LMC influence the frequency and amplitude of kinematic signatures in the outer disk respectively.

\subsection{Tidal features in the $R$--$V_\phi$ Plane}

In this section, we look for radial and vertical velocity signatures of the tidally induced spiral arm in the $R$--$V_\phi$ plane \citep{Ramos2018, Khanna_2019,Antoja2021,Bernet_Ramos2022, Bernet2022,2023Xu, 2024Cao}. Our sample reaches farther distances than previous studies, such as \cite{Wheeler2022} and \cite{Antoja2021}, so our goal is to verify that we see the $-V_R$ and $-V_Z$ regions of the MRi overdensity associated with a tidally induced spiral arm in $R$--$V_\phi$ space. 

The left panel of Figure \ref{fig:R_V_phi_VR_VZ} shows the number density distribution in $R$--$V_\phi$ with two bins per kiloparsec in $R$ and 0.25 bins per kilometers per second in $V_\phi$. We plot only our MRi region selection from the MSTO anticenter sample ($0<Z<5$ kpc) at $Y>0$ kpc to focus on the region where we expect $-V_R$ kinematics associated with the MRi overdensity (see Figure \ref{fig:XY_vr_vphi_vz}).
The median $V_R$ and $V_Z$ of the stars are color-coded in the middle and right panels of Figure \ref{fig:R_V_phi_VR_VZ}, respectively. For all panels, we remove bins with less than 20 stars, and show density contours in black. The white circles show the median $V_\phi$ at each $R$. 
The middle and right panels of Figure \ref{fig:R_V_phi_VR_VZ} demonstrate 
that the region of stars with the $-V_R$ and $-V_Z$ motions associated with the MRi overdensity  also have the median $V_\phi$ at each $R$, indicated by the white circles overlapping with the $-V_R$ and $-V_Z$ region between $14<R<18$ kpc. This shows that the stars that are in the MRi overdensity, between $14<R<18$ kpc (left panel), are corotating with the rest of the disk (i.e., moving at the median $V_\phi$) and also have $-V_R$ (middle panel). This is similar to the third panel of Figure \ref{fig:XY_vr_vphi_vz}. 

To summarize, the ridges in $R$--$V_\phi$ space show distinct patterns of $-V_R$ and $-V_Z$. Since the MRi overdensity has median $-V_R$ and $-V_Z$ kinematics (Section \ref{sec:MRi}), we associate those stars in the $R$--$V_\phi$ ridges with the MRi overdensity.

\begin{figure}[h!]
    \centering
    \includegraphics[width=1.0\linewidth]{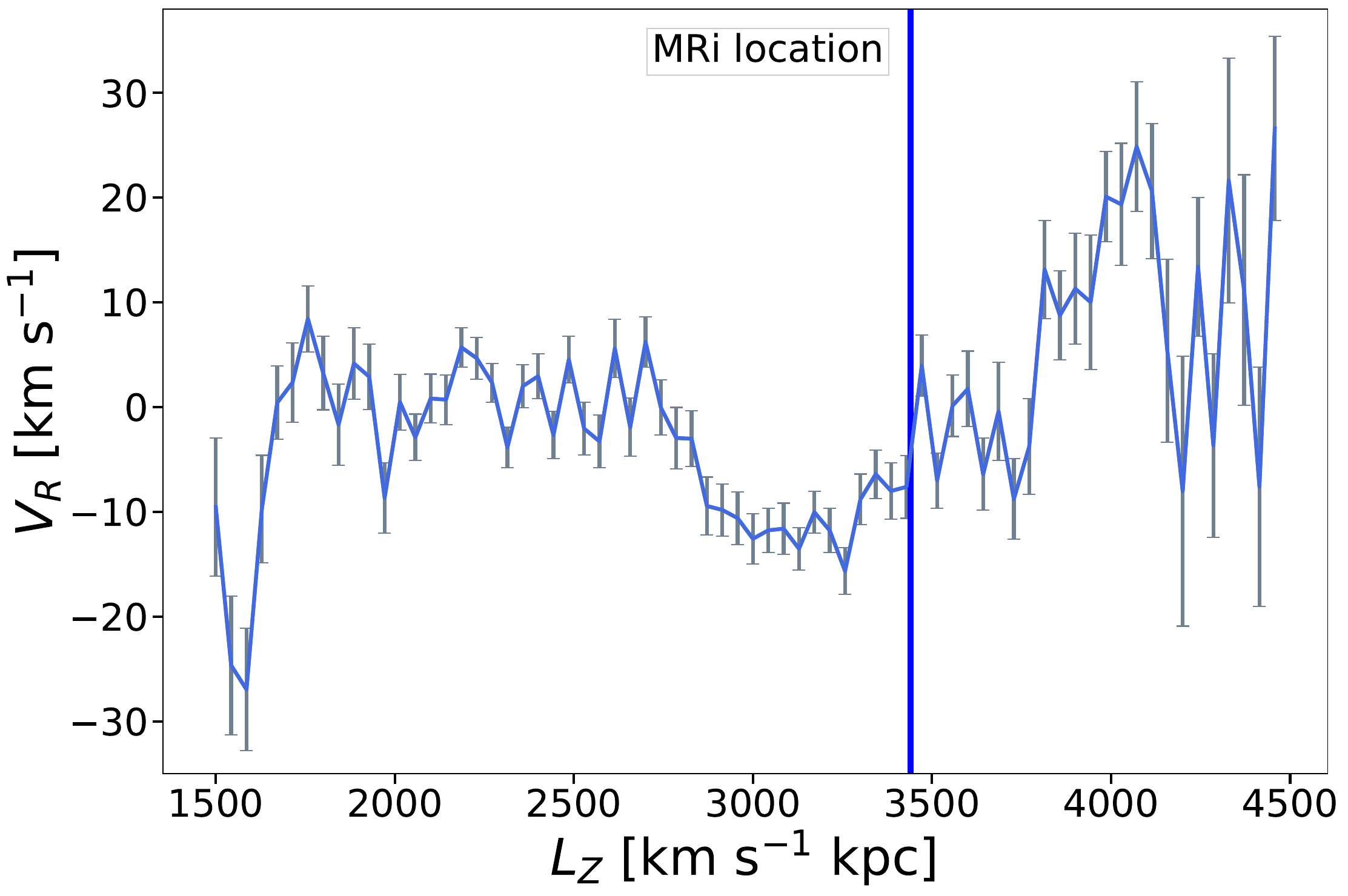}
    \caption{The mean $V_R$ in a given bin in a range in $L_Z$ of 1400 to 4500 km s$^{-1}$ kpc over 70 bins in $L_Z$. The error bars represent the standard error on the mean. The blue vertical line indicates the $L_Z$ associated with the center of the MRi region, $R=16$ kpc.}  
    \label{fig:VR_Lz}
\end{figure}

\subsection{Sgr Pericenter Timing Analysis}
\label{sec:timing}

For a tidally induced spiral arm,
the number of wraps increases in frequency as a function of time \citep[e.g.][]{2022Antoja}. As described in Section \ref{sec:MRi}, the location of minima in $V_R$ correspond to the overdensity associated with the spiral arm as seen in Figure \ref{fig:XY_vr_vphi_vz}. As these arms wrap up over time, they create an oscillatory pattern in $V_R$ versus $R$ (see Figure 2 in \citealt{2022Antoja}). Therefore, if we know the rate at which the arms wind up, the frequency of oscillations in $R$ tells us how long the arms have been winding up. The time it takes the spiral arms to wind up depends on the pattern speed of the disk, which, for corotating tidally induced spiral arms, is the circular speed ($V_c$).

We follow the analysis described in \cite{2022Antoja} to measure the winding frequency and relate that to the pericenter passage time. 
We use a Fourier analysis of the observed $V_R$ oscillation pattern to recover the time of recent pericenter passages of external perturbers of the MW, like Sgr.  

Following Section 4 of \cite{2022Antoja}, we use a one-dimensional analysis to measure the frequency of the oscillations in $V_R$ in the anticenter region of the MW disk. A two-dimensional frequency analysis would require more complete coverage of the disk compared to our limited range in Galactic longitude \citep{2022Antoja}.

We measure the frequency of oscillations in $V_R$ versus $L_Z^{\frac{n-1}{n+1}}$ where $n$ is the MW circular velocity slope and $L_Z$ is the angular momentum defined as $L_Z= R \times V_\phi$. For a circular velocity curve of the form $V_c$($R$) $= V_0$($\frac{R}{R_0}$)$^n$, using $L_Z^{\frac{n-1}{n+1}}$ as the independent variable removes the dependence of the spiral arm winding rate on $R$. We also assume that the orbits of the stars in the plane of the disk can be described by the epicycle approximation. These result in a relationship between the frequency of the $V_R$ versus $L_Z^{\frac{n-1}{n+1}}$ oscillations and the time the arms have been winding since pericenter passages of the perturber. See also the discussion of Equation 7 in \cite{2022Antoja}. We adopt a flat circular velocity curve for the MW \citep{2023Poder, 2024Binney_vasiliev}, 
$n=0$, which corresponds to $L_Z^{-1}$. We explore the impact of this choice later in this section. 
For this analysis, we select the MRi region MSTO anticenter sample with an $R$ range of $8<R<23$ kpc, and $Z<5$ kpc, as used in Section \ref{sec:MRi}, and an $l$ range of $175^\circ <l<185^\circ$.

Figure \ref{fig:VR_Lz} shows the median $V_R$ in bins of $L_Z$ (similar to the right-most panel in Figure 24 in \citealt{2022Antoja}). We compute the median $V_R$ in 75 bins over a range in $L_Z$ between 2000 to 4500 km s$^{-1}$ kpc. 
The error bars represent the standard error on the mean $V_R$. The blue vertical line indicates the $L_Z$ associated with the center of the MRi region, $R=16$ kpc, which corresponds to the location of the $-V_R$ dip.

We compute a Fast Fourier Transform (FFT) to measure the frequency of $V_R$ in $L_Z^{\frac{n-1}{n+1}}$. We can then relate the peak frequencies to the pericenter time as described below.

\begin{figure}
    \centering
    \includegraphics[width=1.0\linewidth]{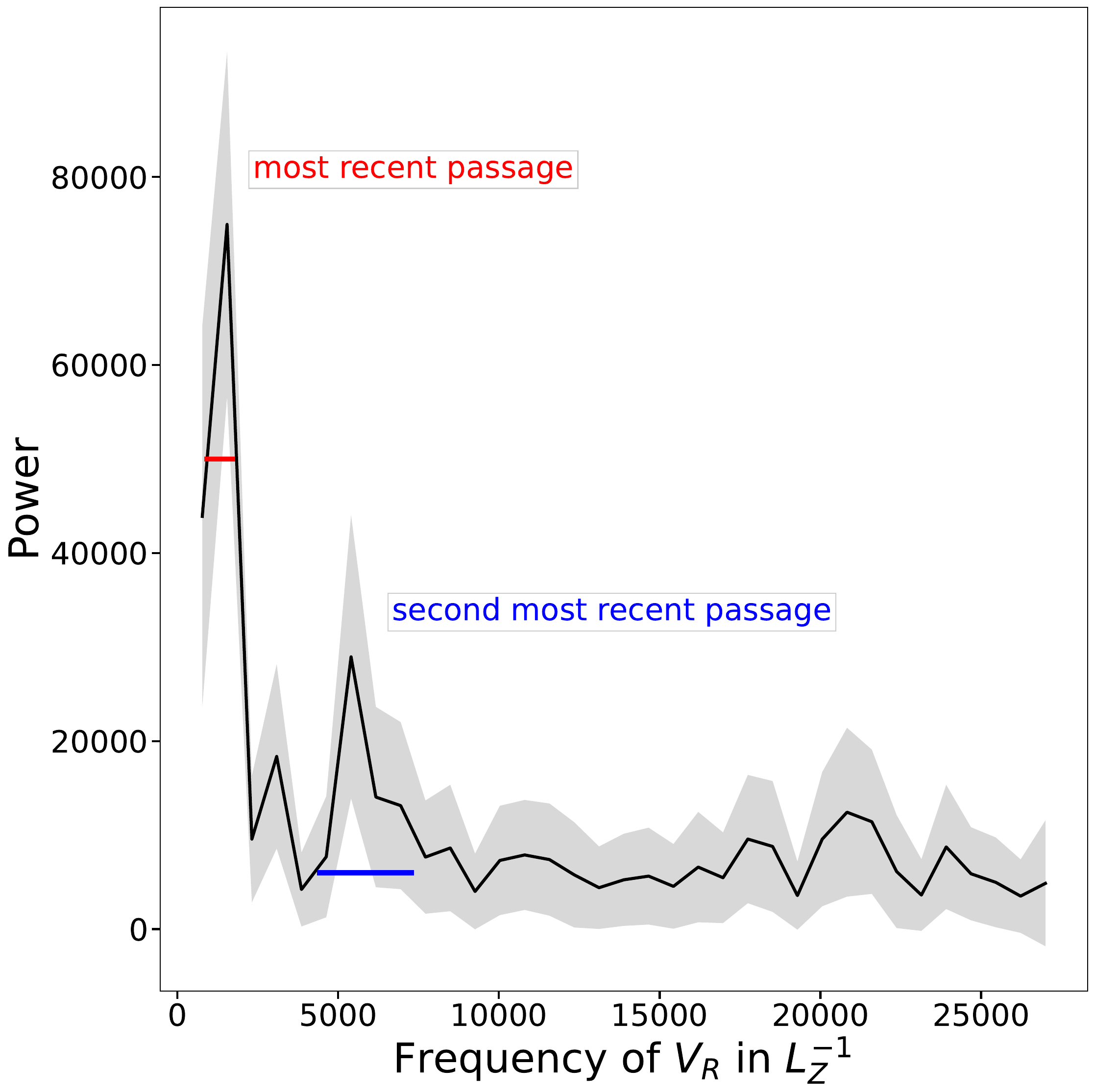}
    \caption{Power spectrum of the Fourier analysis calculated from the data shown in Figure \ref{fig:VR_Lz}. The black line represents the median power spectrum after Monte-Carlo sampling the $V_R$ uncertainties from Figure \ref{fig:VR_Lz}. The 1$\sigma$ region is indicated in gray. The red and blue horizontal lines indicate the uncertainty in the two significant peaks calculated by fitting a Gaussian to each peak. The mean and $1\sigma$ values from each fitted peak are 1313.0 $\pm$ 477.2 ((km s$^{-1}$ kpc)$^{-1})^{-1}$ and 5878.6 $\pm$ 1471.9 ((km s$^{-1}$ kpc)$^{-1})^{-1}$, respectively.} 
    \label{fig:fft}
\end{figure}

We perform Monte-Carlo sampling using the uncertainties on $V_R$, run the FFT 1,000 times, and plot the average power spectrum as the black line and 1$\sigma$ as the gray region in Figure \ref{fig:fft}. There are two clear peaks in frequency, and further below, we show how these peaks are related to the two most recent pericenter passages of Sgr.
We calculate the width of each peak by fitting a Gaussian to each peak. We use the widths to estimate the uncertainties on the locations of the peaks. The red and blue horizontal lines indicate the width of the two significant peaks. The corresponding average frequency and $1\sigma$ values of the peaks are 1313.0 $\pm$ 477.2 ((km s$^{-1}$ kpc)$^{-1})^{-1}$ (red) and 5878.6 $\pm$ 1471.9 ((km s$^{-1}$ kpc)$^{-1})^{-1}$ (blue). The units of frequency are 1/($L_Z^{\frac{n-1}{n+1}}$). We leave the units in an unsimplified form to emphasize that the values of frequency are dependent on the assumed circular velocity curve slope.
We compare our frequencies to \cite{2022Antoja} in Table \ref{tab:time_table}, and find they are consistent within the uncertainties.

Equation \ref{time_eq} (equivalent to Equation 11 of \cite{2022Antoja}) relates the peak frequency to the winding time, and therefore the time since the pericenter passage of the perturber.

\begin{equation}
    \Delta L_Z ^{\frac{n-1}{n+1}} = (\frac{V_0}{R_0^n})^{\frac{-2}{1+n}}\frac{1}{1-0.5\sqrt{2(n+1)}}\frac{\pi}{t}
\label{time_eq}
\end{equation}
In Equation \ref{time_eq}, $t$ is the pericenter passage time we are solving for, $\Delta L_Z^{\frac{n-1}{n+1}}$ is the location of the peak we recover in Figure \ref{fig:fft}, $n$ is the slope of the circular velocity curve, $V_0$ is the azimuthal velocity of the disk at the location of the Sun, $R_0$ is the distance of the Sun from the center of the Galaxy.
As before, we use $n=0$, $V_0$= 239.26 km s$^{-1}$ and $R_0$ = 8.277 kpc \citep{ 2020Reid, 2022GRAVITY, 2023Poder, 2024Binney_vasiliev}.

We calculate the times of the pericenter passages corresponding to the two peaks in Figure \ref{fig:fft} are 0.25 $\pm$ 0.09 Gyrs and 1.10 $\pm$ 0.28 Gyrs from present day. In Table \ref{tab:time_table}, we compare our results to previous studies calculating the pericenter times of Sgr using simulations or other observational methods, and we find our results are consistent with most studies within our uncertainties.

Here, we investigate the impact of our assumption of a flat circular velocity curve for the MW. From recent measurements in the MW \citep{2019Eilers, 2015Bovy, 2020Sofue, 2017McMillan}, we choose slopes of $n=-0.01$, $-0.5$, and $-0.1$. Repeating the same procedure using Equation \ref{time_eq} and inserting these different slopes, we recover two significant peaks for all slopes, and find that steeper slopes are correlated with more recent passage times. All of the times that we find from the different slopes are within $1\sigma$ of pericenter passage time using the flat circular velocity curve
except for the highest frequency peak for the $n=-0.1$, which is $1.10\sigma$ different.

In this section, we connect the kinematics of the MRi region to the accretion history of Sgr. Using the model for tidally induced spiral arms from \cite{2022Antoja}, we relate the frequency of $-V_R$ regions to the approximate time from which the arms have been winding up due to Sgr's impact. Our methodology gives us an estimate of the two most recent pericenter passage times of Sgr, 0.25 $\pm$ 0.09 Gyrs and 1.10 $\pm$ 0.28 Gyrs from present day. Comparing our analysis to other methods of estimating the pericenter passage timing of Sgr, we find our results are consistent.

\begin{deluxetable*}{ccccc}
\tablecaption{Summary of the two most recent pericenter passage times of Sgr in Gyrs.}
\label{tab:time_table}
\tablewidth{0.5pt}
\tablehead{
\colhead{Literature\tablenotemark{a}}&\colhead{Method}&\colhead{$\Delta L_Z$ \tablenotemark{b}}&\colhead{most recent  passage time}&\colhead{second passage time}}
\startdata
~~LM10&Sgr orbit modeling&...&0&1.3\\
~~dlV15&Sgr orbit modeling&...&...&1.1\\
~~DL17&Sgr orbit modeling&...&0&1.3\\
~~L19&Modeling Phase Spiral&...&0&0.5\\
~~RL20&Local Star Formation Rate&...&0.075&1.0\\
~~A22&Frequency Analysis&$<$3500,  7100-11800&$<$0.6&1.3-2.1\\
~~\textbf{This work}&Frequency Analysis&1313.0 $\pm$ 477.2, 5878.6 $\pm$ 1471.9 & 0.25 $\pm$ 0.09&1.10 $\pm$ 0.23\\
\enddata
\tablenotetext{a}{LM10 is \cite{2010Law}, ldV15 is \cite{2015delaVega}, DL17 is \cite{2017Dierickx}, L19 is \cite{Laporte2019}, RL20 is \cite{2020Ruiz_lara_nature}, A22 is \cite{2022Antoja}.} \tablenotetext{b}{Only the frequency analysis method calculates the $\Delta L_Z$ using kinematics of tidally induced spiral arms.}
\end{deluxetable*}

\begin{figure}
    \centering
    \includegraphics[width=1\linewidth]{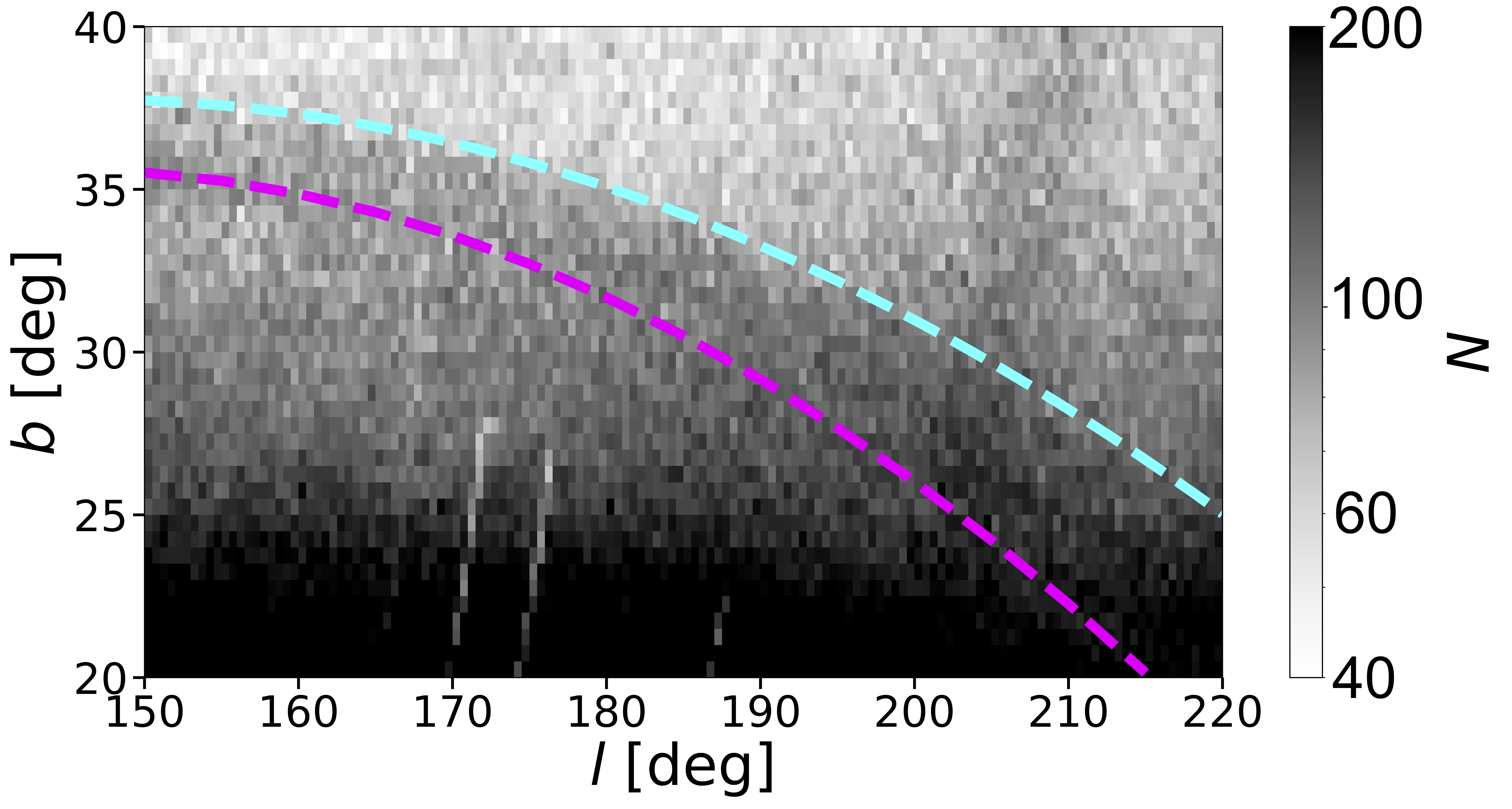}
    \caption{2D histogram of the ACS region in $l$, $b$ space using two bins per degree along each axis of stars from Pan-STARRS photometry with $0.2<g-r<0.4$ and $18<g<21$ mag. We fit a parabola to the sharp upper edge of the ACS overdensity in $b$, indicated by the cyan dashed line. We fit a parabola to the peak overdensity in $b$ which is illustrated by the magenta dashed line. We do not fit a lower edge in $b$ of the ACS overdensity due to challenges with disk contamination at lower Galactic latitudes. The vertical overdensity present on the right-hand side of the figure is the Sgr stream.} 
    \label{fig:l_b_panstarrs}
\end{figure}

\begin{figure*}
    \centering
    \includegraphics[width=1.0\linewidth]
    {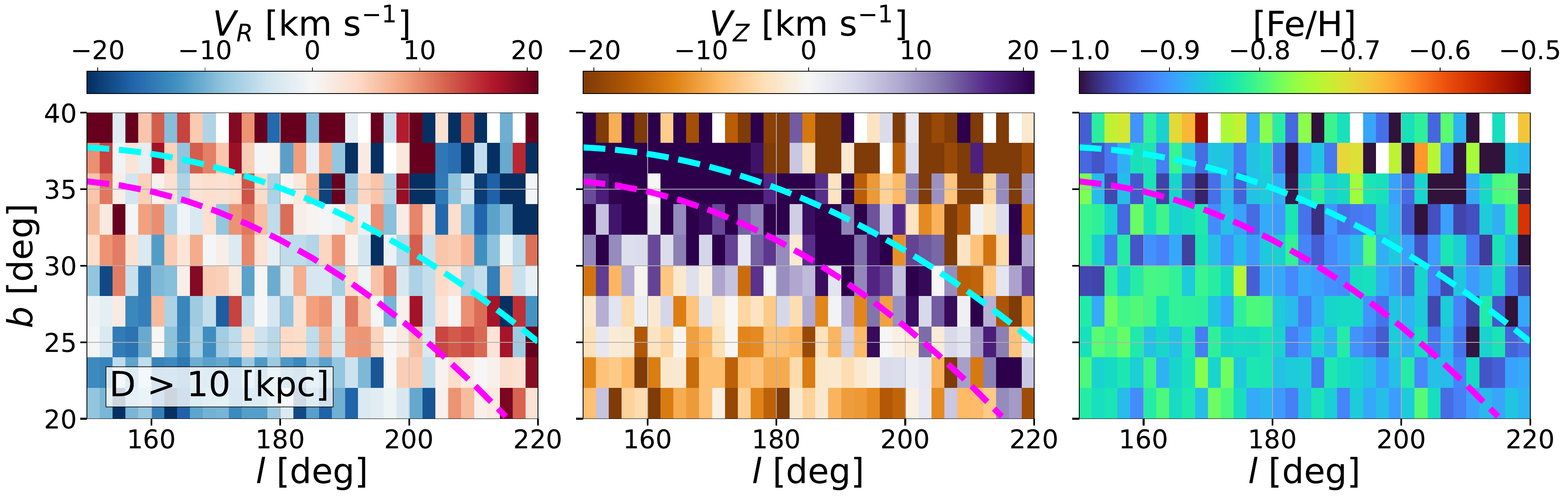}
    \caption{Left: MSTO anticenter sample with distance from the Sun greater than 10 kpc plotted in $l$, $b$ space with bins color mapped to median $V_R$, and has two degrees per bin in both $l$ and $b$. Middle: Same as the left panel, but bins are color-mapped to median $V_Z$. Right: Same as the left panel, but bins are color-mapped to median metallicity. Compared to the top row in Figure \ref{fig:lb_RZ}, applying the distance cut allows us to see the $-V_R$ and $-V_Z$ region extend to lower latitudes and higher longitudes. The cyan line in all the panels is a parabolic fit that indicates the upper edge of the photometric overdensity of the ACS. We see the cyan line also traces the upper edge of the median $-V_R$, $-V_Z$, and slightly more metal-poor region. The magenta line is a parabolic fit indicating the peak photometric overdensity of the ACS. We see the region of median $-V_R$ and $-V_Z$ reaches latitudes lower than the megenta line.}
    \label{fig:l_b_acs}
\end{figure*}

\subsection{The ACS Overdensity}
\label{kine_ACS}

\begin{figure*}
    \centering
    \includegraphics[width=1.0\linewidth]{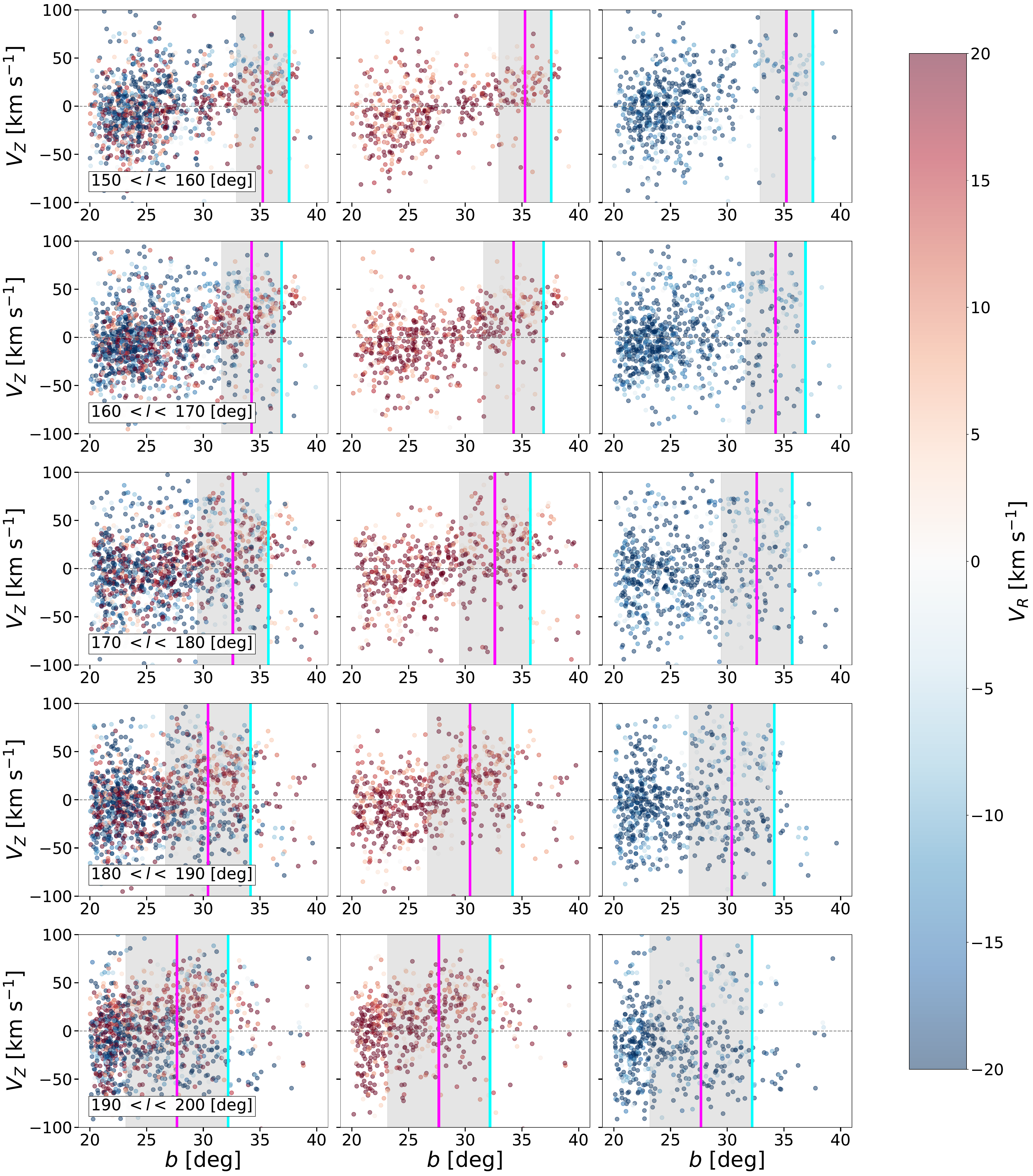}
    \caption{MSTO anticenter sample with heliocentric distance $>10$ kpc plotted in $b$--$V_Z$. Left column plots all stars color coded to $V_R$ in $l$ bins of 10$^\circ$ from $150^\circ<l<200^\circ$ as labeled in each row. The middle column plots only stars with $+V_R$, while the right column plots stars with $-V_R$. The cyan line represents the $b$ value associated with the parabolic fit to the upper edge of the ACS photometric overdensity. The magenta line shows the $b$ value associated with the parabolic fit to the peak overdensity of the ACS photometric overdensity. The gray region indicates the overdensity's width centered on the peak density (magenta line), assuming the Galactic latitude distance between the upper edge of the overdensity and the peak density is the same as the lower edge of the overdensity and the peak density. We find a positive correlation between $b$ and $V_Z$ for stars with $+V_R$ that extends to lower latitudes than the ACS photometric region, and is not just confined to the overdensity region. These velocities, $-V_R$ and $-V_Z$, especially seen in the right bottom two panels, are associated with the MRi overdensity.}
    \label{fig:b_VZ_all}
\end{figure*}

We base our definition of the ACS region on the location where we visually see the ACS in our MSTO anticenter sample, as well as a photometric sample. Using Pan-STARRS photometry \citep{2016ChambersSTARRS}, we select stars from that survey with $0.2<g-r<0.4$ and $18<g<21$ mag. We show this photometric selection as a two-dimensional histogram in Figure \ref{fig:l_b_panstarrs} in two bins per degree along $l$ and $b$. This figure displays the region in $150^\circ<l<220^\circ$ and $20^\circ<b<40^\circ$ which is the same as our DESI spectroscopic MSTO anticenter sample.

We define the ACS overdensity to be stars within $150^\circ<l<200^\circ$. This $l$ cut avoids the photometric contamination of the Sgr stream at $l>200^\circ$, as seen as a vertical overdensity in Figure \ref{fig:l_b_panstarrs}.
The ACS overdensity shows a clear upper edge in $b$ and peak density in our Pan-STARRS sample. The cyan line is a parabolic fit to the upper edge in $b$ of the ACS overdensity. The magenta line is a parabolic fit to the peak density in $b$ of the ACS region. We do not fit for the lower edge of the ACS overdensity because, at lower $b$, disk stars begin to dominate. 
Based on these two parabolas, we define a region in the sky where the ACS overdensity is present. We also notice that the ACS overdensity broadens as it gets closer to the disk at larger $l$ and lower $b$.

When analyzing the ACS region in our MSTO anticenter sample, we define the region in $l$, $b$ to be the same as the photometric sample in Figure \ref{fig:l_b_panstarrs}, as well as a minimum heliocentric distance of 10 kpc.
We apply a distance cut rather than a $Z$ cut for the ACS region, so we can follow stars associated with the ACS overdensity to low $Z$. With a distance cut $>10$ kpc, the lowest $Z$ value in our MSTO anticenter sample is 3.5 kpc. This means, based on our definition of the MRi region, we expect to find some contamination from the MRi overdensity at lower latitudes ($b\lesssim25^\circ$).

\subsection{Kinematics of the ACS Region}
\label{kine_ACS}

\subsubsection{Kinematics in Observational Coordinates $l$, $b$}
In Figure \ref{fig:l_b_acs}, we plot our DESI MWS MSTO anticenter sample of the ACS region described in the previous section in $l$, $b$ space. The same parabolas fitted to the photometric sample are shown in each panel, and all the panels have two degrees per bin in both $l$ and $b$. In the left panel, color coded to the median $V_R$, we see $+V_R$ in the region between the two parabolas, and a small region below the magenta line. The majority of bins at lower latitudes with $-V_R$ are associated with the MRi region. The middle panel is color coded to the median $V_Z$, and, similar to the left panel, we see the region between the two parabolas mostly encapsulates the $+V_Z$ region on the sky, where a small region below the magenta line also shows $+V_Z$ values. The part of the sky at lower latitudes and appears to have mostly $-V_Z$ velocities is associated with the MRi region. Lastly, the right panel is color-mapped to the median metallicity, and we see that the region between the two parabolas has on average slightly lower metallicity.
We refer the reader back to the right panel of Figure \ref{fig:lb_RZ} for our results of the metallicities of the region.

Figure \ref{fig:l_b_acs} demonstrates that the median kinematics in the ACS overdensity are different from the median kinematics at lower latitudes. 
In order to test whether the ACS overdensity is a discrete stream-like object versus part of a broader structure in the disk, we need to identify whether the kinematics of the ACS overdensity continue to lower latitudes. From Figure \ref{fig:l_b_acs}, we cannot say whether the $+V_R$ and $+V_Z$ kinematics are obscured by lower latitude, disk-like stars.

\subsubsection{The ACS as Part of a Broader Disk Structure}
To determine whether the kinematics associated with the ACS overdensity are continuous or not at lower latitudes, we plot $V_Z$ versus $b$ at different $l$ bins shown in Figure \ref{fig:b_VZ_all}. In the left column, between $150^\circ<l<200^\circ$, we plot all the stars in the MSTO anticenter sample with heliocentric distance $>10$ kpc with $l$ bins of 10$^\circ$ for each row. The stars are color coded to their $V_R$ value. The middle column only plots stars with $+V_R$, and the right column only plots stars with $-V_R$, still color coded by $V_R$.
The cyan and magenta vertical lines indicate the $b$ value of the upper edge parabola and peak overdensity parabola, respectively, for each row. The gray region indicates the symmetric width about the peak density, assuming the latitude distance between the upper edge of the overdensity and the peak density is the same as the lower edge of the overdensity and the peak density. As mentioned previously, the ACS photometric overdensity broadens at larger $l$ (i.e., where the overdensity trends to lower $b$. This is reflected in the size of the gray region in the different rows as well.

In the middle column of Figure \ref{fig:b_VZ_all}, the $+V_R$ stars extend to the upper edge of the photometric overdensity, but are not concentrated in the overdensity, as they continue to lower $b$. In the right column, the majority of the $-V_R$ stars do not extend up to the overdensity, and are mainly concentrated to lower $b$. We find that, for the $+V_R$ stars, there appears to be a positive correlation in $V_Z$ as a function of $b$ seen in all the rows. In other words, at higher $b$, the stars have more positive $V_Z$. From this column, it is important to note that, in all $l$ bins, the positive correlation is present at the upper edge of the overdensity and continues to lower latitudes, past the nominal ACS overdensity, and connects to the lower latitude disk stars. 
The gap at $b\sim27^\circ$ in the top panel may be due to the incompleteness of our data rather than a physical gap; therefore, we cannot make any claims about density discontinuity between the ACS overdensity and lower latitude stars.

The majority of the $-V_R$ stars in the right column of Figure \ref{fig:b_VZ_all} appear to be located at lower $b$ than at the photometric overdensity marked by the gray region.
There is a slight preference for $-V_R$ stars to also have $-V_Z$ velocities as seen in the bottom two rows. The slight correlation is most likely due to contamination from the MRi overdensity at these lower Galactic latitudes. This is in agreement with what we found in Section \ref{sec:MRi} where the MRi overdensity is associated with mostly $-V_R$ and $-V_Z$ velocities.

\begin{figure*}
    \centering
    \includegraphics[width=0.9\linewidth]{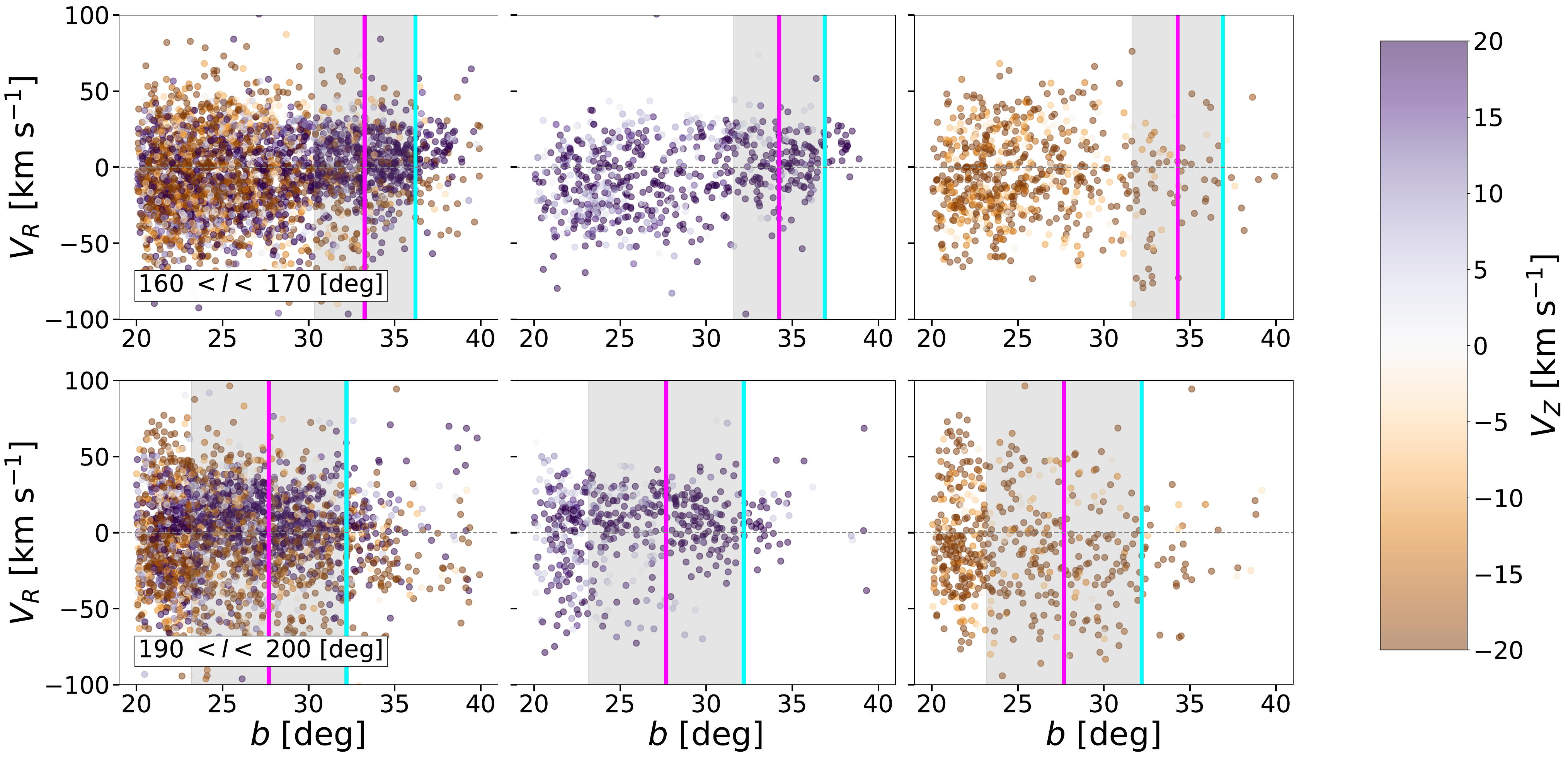}
    \caption{MSTO anticenter sample with distance from the Sun greater than 10 kpc in $b$--$V_R$ space. Left column plots all stars color coded to $V_Z$, middle column plots stars with $+V_Z$, and right column plots stars with $-V_Z$. The top row plots stars between  $160^\circ<l<170^\circ$, and the bottom row plots stars between $190^\circ<l<200^\circ$ as labeled in each row. The cyan line, magenta line, and gray region are the same as Figure \ref{fig:b_VZ_all}. Similar to Figure \ref{fig:b_VZ_all}, we find the $+V_Z$ kinematics associated with the overdensity follow the upper edge of the overdensity and extend to lower $b$. We find the $-V_Z$ kinematics are concentrated at lower $b$. There is a slight preference for $-V_Z$ stars to also have $-V_R$ velocities. Again, the $-V_R$ and $-V_Z$ velocities at low Galactic latitudes are associated with the MRi overdensity.}
    \label{fig:b_VR_small}
\end{figure*}

To go along with the $b$--$V_Z$ panels, Figure \ref{fig:b_VR_small} shows panels of $b$--$V_R$ in a $l$ bin of $160^\circ<l<170^\circ$, where the ACS photometric overdensity is at its highest $b$ in our sample, and a $l$ bin of $190^\circ<l<200^\circ$, where the overdensity is at a lower $b$ and is broader. The cyan line, magenta line, and gray region are the same as in Figure \ref{fig:b_VZ_all}. The stars are color coded to $V_Z$ with the left column plotting all stars, the middle column plotting $+V_Z$ stars, and the right column plotting $-V_Z$ stars. Similar to Figure \ref{fig:b_VZ_all}, we see that the $+V_Z$ stars are dominant at the ACS photometric overdensity but also extend to lower latitudes. The $-V_Z$ stars primarily lie at lower latitudes. Neither $+V_Z$ nor $-V_Z$ show any correlation between $b$ and $V_R$. We do see a slight preference for $-V_Z$ to also have $-V_R$, similar to Figure \ref{fig:b_VZ_all}, indicating we have some contamination at lower latitudes from the MRi overdensity.

To summarize, the extent in Galactic latitude of the $+V_R$ and $+V_Z$ kinematics trace the upper edge of the photometric overdensity of the ACS. However, the kinematic signatures we find in the ACS photometric overdensity ($+V_R$ and $+V_Z$) are not confined to that overdense region and extend to lower Galactic latitudes. The ACS overdensity is not a kinematically discrete group of stars separate from the rest of the disk.
The $+V_R$ and $+V_Z$ regions at high $b$ in Figure \ref{fig:l_b_acs} masqueraded as a discrete, stream-like structure because of the large, co-located  population of stars with $-V_R$ and $-V_Z$ kinematics dominating at lower latitudes.
Those $-V_R$ and $-V_Z$ motions are associated with the MRi overdensity.

\section{Discussion}
\label{sec:Discussion}

Here we discuss the results from Section \ref{sec:results} in context with other recent literature. In Section \ref{sec:TISA_discuss}, we comment on well-known spiral arms in the MW compared to the outer disk overdensities, other simulations that can produce outer disk overdensities from interactions between the MW and Sgr, the effect of the LMC on the outer disk, and previous work relating tidally induced spiral arms to the $Z$--$V_Z$ phase spiral. In Section \ref{sec:R_Vphi_discuss}, we discuss other explanations for ridges in $R$--$V_\phi$ space including the influence of resonant orbits from the bar on ridges out to $R \sim 12.4$ kpc. In Section \ref{sec:timing_discuss}, we expand on the Sgr pericenter passage timing comparison with the  literature we compare to in Table \ref{tab:time_table}. In Section \ref{sec:ACS_discuss}, we discuss previous interpretations of the ACS overdensity and simulations of low Galactic latitude streams of disk stars from \cite{Laporte2019_feathers}. We also comment on other overdensities found in the anticenter region and their possible connection to the ACS and MRi.

\subsection{Tidally Induced Spiral Arms in the Context of Disk Substructure}
\label{sec:TISA_discuss}

The two leading theories of the mechanisms driving spiral arm formation in galaxies are the quasi-stationary density wave model \citep{1964LinShu} and the dynamic arm scenario, caused by gravitational instabilities or interactions with external perturbers \citep[Chapter 6 of][]{BinneyTremaine2008, 1985carlberg}. 
Recent studies show that dynamic spiral arms are a plausible explanation for spiral structure near the solar neighborhood, like the Orion spur and the Perseus arm, using observations such as the lack of age gradients within the arms \citep{2025Liu}.  Studies also find that for dynamic spiral arms, the sign of $V_R$ flips around the center of the arm; when the velocities are toward the arm, the arm is growing, and when the velocities are away from the arm, it is dissipating \citep{2024Funakoshi, 2015Grand}. According to the simulations, these dynamic spiral arms grow from a small density perturbation and then disrupt \cite{2015Grand}. In the case of the Perseus arm, the observed stars on the inside edge of the arm (smaller guiding center radius) have $V_R$ pointing inward, whereas the observed stars on the outer edge of the arm have $V_R$ pointing outward \citep{2024Funakoshi}. So they interpret this as the Perseus arm disrupting \citep{2024Funakoshi}. 
It is also important to note that we are not suggesting the MRi is part of the Norma-Outer arm, which is located at 10 $<R<$ 15 kpc \citep{2019Reid}. The MRi is an overdensity that resides beyond 15 kpc and is mostly made up of older, more metal-poor stars, whereas the well-known arms are gas-rich and star-forming.

Previous N-body simulations of the interactions between the MW and Sgr have been able to create MRi-like overdensities with similar kinematics (minima in $V_R$) and hypothesize the MRi to be an extended spiral arm \citep{2011Purcell, 2016gomez}. Many studies have found the $-V_R$ signature associated with the MRi \citep{2020Laporte, 2024Qiao, Borbolato2024}. Others have found the $V_\phi$ motion of the MRi overdensity to be similar to that of the thin disk \citep{2018Deason, Borbolato2024}. However, we know of no other observational study connecting the $-V_R$ motion and inflection in $V_\phi - <V_\phi$($R$)$>$ of the MRi to a co-rotating spiral arm pattern.

Another important point to consider is the longevity of these types of disk structures. \cite{Struck_2011} and \cite{2018bLaporte} have demonstrated, via N-body simulations, that the spiral arms and feathers in the disk created by satellite interactions could last for a few Gyrs, meaning it is plausible that we observe a tidally induced spiral arm a few Gyrs after its formation.

We can rule out the warp or flare of the disk as the dominant factor in creating the overdensity seen in Figure \ref{fig:XY_vr_vphi_vz}. According to studies like \cite{2019Li_flareandwarpgaiadr2}, \cite{2020Xu}, and \cite{2023Han}, the warp reaches $Z\sim1.5$ kpc at $R=20$ kpc. However, geometrically, the warp is not at the correct longitude to create an apparent overdensity at the location of the MRi. The line-of-node is approximately through the Galactic center--solar neighborhood and continues through to the Galactic anticenter \citep{2006Momany}.
Futhermore, the flare of the disk reaches a scale height of $Z \sim2.5$ kpc by $R\sim20 $ kpc \citep{2019Thomas, 2019Li_flareandwarpgaiadr2, 2022chrobakova, 2024Uppal}.
The flare demonstrates how disk stars can be found at these heights, but neither the warp nor the flare alone can explain the kinematic signatures we see.

Several previous works are in agreement that the MRi overdensity is likely the result of the gravitational influence of a recent satellite interaction.
\cite{2016gomez} uses cosmological simulations to show, based on the vertical height of the MRi, that the origin may be due to a satellite flyby and could result in a leading spiral arm.
\cite{2011MichelDansac} also used idealized simulations to show how the passages of Sgr could create such an overdensity. 
In \cite{2018bLaporte}, they argue that Sgr is the main culprit for current perturbations and corrugations in the MW disk using N-body simulations of the orbital history of Sgr and the LMC. Our work agrees with their claims that the MRi can be created by the tidal forces from the accretion of the Sgr. Furthermore, we connect the MRi to regions of $-V_R$ and inflection in $V_\phi - <V_\phi$($R$)$>$ which match the kinematics of a tidally induced spiral arm.

Another possible scenario is that the LMC system may have some influence on perturbing the disk. The mass of the LMC at infall \citep[2 x 10$^{11} M_\odot$;][]{2012Besla} is large enough to have some gravitational impact on the disk \citep{2021Petersen}.
Simulations from \cite{2018bLaporte} show that the LMC does not add any new corrugations to the MW disk but does produce large-scale features like warps and flares.
\cite{2024Stelea} models the interaction of the LMC and Sgr with the MW separately and together to compare their impact on the structure and kinematics of the disk. They also find that the effect of the LMC can change the amplitude of the MRi-like features caused by Sgr.
The inclusion of the LMC along with Sgr in their simulations qualitatively matches our analysis of the MRi. \cite{2024Stelea} also finds a slight offset in azimuthal location between the location of their $-V_R$ region and $-V_Z$ region which may be related to the strong $-V_Z$ region appearing to trail towards negative $Y$ in our observations of Figure \ref{fig:XY_vr_vphi_vz}.

\cite{2024Stelea} also find that with a cold disk model, secular instabilities are more visible in the inner region. However, those instabilities are not large enough to change the sign of $V_R$ or $V_Z$, but do enhance the amplitude, and the instabilities do not appear to affect the outer disk, where MRi is located.

Numerical simulations by \cite{2025Asano} and \cite{Laporte2019} demonstrate a connection between the tidally induced spiral arms and the phase spiral. The N-body simulation from \cite{2025Asano} shows that tidally induced spiral arms from Sgr excite two-arm phase spirals in the MW disk.
In Section 6.2 of \cite{Laporte2019}, they assert that it is possible for stellar overdensities in the outer disk to appear as disconnected ``chunks'' of the spiral pattern in $Z$--$V_Z$. 
Based on our analysis, we cannot claim the tidally induced spiral arm signature of the MRi region is related to the $Z$--$V_Z$ phase spiral. Future work should test these predictions with spectroscopic observations closer to the plane of the disk in the outer regions of the disk.

\subsection{Transient Spiral Structure in $R$--$V_\phi$}
\label{sec:R_Vphi_discuss}

Our association of the $-V_R$ ridge in the $R$--$V_\phi$ plane with the MRi (Figure \ref{fig:R_V_phi_VR_VZ}) and a tidally induced spiral structure is not the only proposed correlation between these ridges and substructures in the MW disk. $R$--$V_\phi$ ridges near the solar neighborhood have been proposed to be moving groups.
The solar neighborhood ridge ($10<R<12$ kpc) in the left panel of Figure \ref{fig:R_V_phi_VR_VZ} may correspond to previously identified moving groups like Hercules and Sirius \citep{Ramos2018, Bernet_Ramos2022}.

These moving groups have also been examined \citep{Monari2019, Bernet2022, Fragkoudi2019} to determine whether their origins are from the influence of external perturbations or resonance orbits under the influence of the Galactic bar. 
The majority of these studies find that resonance orbits from the Galactic bar are the dominant mechanism within the solar neighborhood.
Studies find the Outer Lindblad resonance (OLR) and corotation resonance locations correspond to moving groups in the solar neighborhood like Hercules, Sirius, and ``hat'', depending on whether the bar's pattern speed is fast, intermediate, or slow, respectively \citep{2017perez_villegas, Wheeler2022, 2024Lucchini}. 
Also, \cite{2021Chiba, 2021Chiba_Schonrich} find that given a decelerating Galactic bar, the Hercules moving group could be stars caught in the corotation resonance orbit.
\cite{2022Clarke} finds that the corotation resonance is located at $6.5<R<7.5$ kpc and the OLR is located at $10.7<R<12.4$ kpc.
It can be argued that the bar's influence cannot exceed its OLR because the quadrupole moment of the bar potential falls off as $R^{-3}$ outside the bar \citep[Sect. 2.4;][]{BinneyTremaine2008}. The MRi is at a much larger distance than these solar neighborhood moving groups, and we therefore find it unlikely that the ridges associated with the MRi in Figure \ref{fig:R_V_phi_VR_VZ} are caused by the bar.

The inner edge of the ridge associated with the MRi kinematics in Figure \ref{fig:R_V_phi_VR_VZ} is located in the same region as the ``AC newridge1'' and ``AC newridge2'' in Figure 16 of \cite{Antoja2021} from their \textit{Gaia} DR3 study. 
We identify the MRi region as the region of $-V_R$ in the middle panel of Figure \ref{fig:R_V_phi_VR_VZ}, whereas  newridge1 and newridge2 from \cite{Antoja2021} are overdensities in star counts in the $R$--$V_\phi$ plane identified using a substructure mask, not kinematics, so we do not make further direct comparison. 

Other studies \citep{Kawata2018, Hunt2018, Khanna_2019, Khoperskov2022} have proposed that the ridges in $R$--$V_\phi$ are associated with spiral arms.
\cite{Khanna_2019} also finds that coupling of radial and vertical oscillations can be generated in isolated disks as well as disks perturbed by a satellite galaxy. These oscillations do not predict the specific correlation of the minimum in $-V_R$ with the overdensity.  

\subsection{Sgr Pericenter Passage Timing Comparison}
\label{sec:timing_discuss}

Our pericenter passage time analysis reveals the two most recent passage times of Sgr, 0.25 $\pm$ 0.09 Gyrs and 1.10 $\pm$ 0.23 Gyrs from present. Table \ref{tab:time_table} compares our results to recent estimates of the Sgr pericenter passage times. Our results are in good agreement with previous timing estimates.

\cite{2010Law} uses an N-body model to simulate the disruption of Sgr in a triaxial MW halo over its orbital history, but they do not include dynamical friction in their analysis. They find that the two most recent closest approaches of Sgr were $\sim$ 0 Gyrs and 1.3 Gyrs ago.
\cite{2015delaVega} uses an orbit integration model from \cite{2014Chakrabarti} of Sgr, assuming an initial mass of Sgr of 10$^{10}$  M$_\odot$ and include the effects of dynamical friction. 
They state the most recent pericenter was 1.1 Gyrs ago, but their orbit shows the current position (at time $\sim$ 0 Gyrs) of Sgr is near its pericenter \citep[see Figure 2;][]{2015delaVega}. 
N-body modeling of the Sgr stream from \cite{2017Dierickx} shows that the period of pericenter passages for the two most recent complete passages is 0 Gyrs and 1.3 Gyrs from present, and it is in good agreement with the dynamic models from \cite{2019Fardal_vandermarel}, which aim to reproduce the Sgr stream features in a simple Galactic potential.
We find that using different methods to calculate the pericenter passage timing of Sgr are in good agreement with one another.

\cite{2011MichelDansac} propose the MRi to be accreted material, which is why we do not include it in Table \ref{tab:time_table}. However, they find that idealized simulations of Sgr show it crossed the disk three distinct times near the current location of the MRi at 0.6, 1.1, and 1.7 Gyrs ago. 
Additionally, \cite{2025Lin} find a two-armed spiral feature in $Z$--$V_Z$ space and estimate perturbation times of $\sim0.32$ Gyrs and $\sim0.5$ Gyrs from present. \cite{2025Lin} suggests the two-armed spiral is a signature of the perturbations originate from two different satellite galaxies.

\subsection{The Anticenter Stream as a Feature of a Vertical Wave}
\label{sec:ACS_discuss}

It is clear from our analysis of the MRi region and ACS region that the $V_R$ and $V_Z$ kinematics of the MRi overdensity are opposite of the ACS overdensity. Similarly, \cite{2020Laporte} finds that the ACS is kinematically decoupled from the MRi.
We also find that the kinematics associated with the ACS overdensity extend to lower Galactic latitudes which demonstrates that the ACS overdensity does not have discrete kinematics that are separate from the disk.

\cite{Laporte2019_feathers} suggests that overdensities like the ACS are tidal tail ``feathers" kicked above the plane of the disk by an external perturber such as Sgr. \cite{Laporte2019_feathers} simulates a MW-like galaxy interacting with a Sgr-like satellite and finds stream-like overdensities in the outer disk, similar to the ACS. They report on the kinematic features of the feathers in observational $l$, $b$ space. 
From their simulations \citep[Section 3.1][]{2018bLaporte}, they detect feathers at the extremities of MRi-like overdensities and find these feathers ubiquitous in their simulation. They find that the shape of these feathers, to an observer, appears as an arc with some vertical peak and, at its edges, smoothly reconnecting to the disk. They also find that the vertical perturbations are excited by interactions with Sgr. 
Figure 9 in \cite{Laporte2019_feathers} finds that feathers spatially resembling the ACS exhibit coherent motion in $Z$ and $R$, with a leading and trailing tail that have opposite kinematics. Our ACS analysis is constrained to the Galactic longitude of our MSTO anticenter sample, whereas the tidal tail feather model from \cite{Laporte2019_feathers} looks at a 360$^\circ$ view of their model. 
Because we only see a limited range of the ACS overdensity, we cannot confirm the full kinematic predictions made by \cite{Laporte2019_feathers}. We also cannot confirm whether tidal tail feathers exhibit kinematics that extend to lower latitudes similar to the ACS overdensity.

In \cite{Laporte2022}, they acknowledge that another interpretation of these observed stream-like overdensities in the anticenter of the disk could be vertical folds in the outer disk \citep{2015Price_Whelan, 2015Xu, 2017li, 2018Sheffield, 2024Cao}. Because our results find the kinematics associated with the ACS overdensity extend to lower Galactic latitudes, we argue that the ACS, rather than a discrete stream-like object, is more likely to be connected to the rest of the disk and is a manifestation of the vertical folds at high Galactic latitudes ($b\sim30^\circ$). The ACS overdensity could be another signature of a disk in disequilibrium, in the middle of phase mixing. 
The vertical folds in the outer disk, including the ACS, could be further evidence for the same kind of vertical disturbance that causes the $Z$--$V_Z$ phase spiral seen in the solar neighborhood \citep{Antoja2018, 2022Antoja, 2024Hunt}. However, that investigation is outside of the scope of this paper. 

Another possible area of exploration to identify what the ACS and MRi could be is to investigate their connection to
the Triangulum-Andromeda \citep[TriAnd;][]{2003Rocha} and/or the Eastern Banded Structure \citep[EBS;][]{2006Grillmair} overdensities. Both TriAnd and EBS are also thought to have formed in situ and exhibit characteristics of MW disk stars \citep{2018Bergemann, 2023Abuchaim}. However, the TriAnd overdensity is located at $-40^\circ < b < -20^\circ$ and $100^\circ < l < 150^\circ$ \citep{2003Rocha} and therefore outside of our sample, so we cannot confirm its possible connection here. The EBS is also primarily outside of the region we analyze, located at $217^\circ < l < 260^\circ$ \citep{2016Morganson, 2018Deason}. 
\cite{2018Deason} argues that the EBS is part of the MRi and marks the edge of the overdensity.
These substructures, MRi, ACS, and TriAnd, can also be connected to the vertical wave in the stellar disk \citep{2015Xu, 2017li, 2018Sheffield}.

\section{Summary}
\label{sec:summary}
In this paper, using MSTO stars in the anticenter region of the MW disk from the first three years of DESI observations, we find the following:

1. The MRi and ACS are separate kinematic disk structures, moving in opposite directions. The MRi has negative radial velocities and negative vertical velocities, while the ACS has positive radial and vertical velocities.

2. The kinematics in the region of the MRi are consistent with those of a tidally induced spiral arm. A tidally induced spiral arm has negative radial velocities and $V_\phi - <V_\phi$($R$)$>$ close to zero at the location of the spiral arm \citep{2022Antoja}.

3. In $R$--$V_\phi$ space, we find an outer ridge made of stars with $-V_R$ values in the region we associate with the MRi. The ridge is most prominently seen in $-V_R$ and $-V_Z$, and corresponds to the median $V_\phi$ per R in a range of $14<R<18$ kpc. This is indicative of corotating motion like that of a tidally induced spiral arm.

4. We perform a Fourier analysis in the region associated with the MRi to determine the two most recent pericenter passages of the Sgr. We find the passages happened 0.25 $\pm$ 0.09 Gyrs and 1.10 $\pm$ 0.23 Gyr from present day. When compared to other estimations of the pericenter passages \citep{2010Law, 2015delaVega, 2017Dierickx, Laporte2019, 2020Ruiz_lara_nature, 2022Antoja}, we are consistent with those values which use a range of different methods to calculate the timing.

5. The kinematics in the region of the ACS are not confined to the overdensity and extend to lower latitudes, connecting with the disk kinematics. The ACS overdensity is further evidence of vertical disequilibrium in the outer disk and could be related to folds of the disk in the anticenter \citep[e.g.][]{2015Xu, Li2017, 2018Sheffield}. 

\begin{acknowledgments}

This material is based upon work supported
by the National Science Foundation Graduate Research Fellowship under Grant No. 2240310.
This work was supported by the U.S. Department of Energy, Office of Science, Office of High Energy Physics, under Award Number DE-SC0010107.
LBeS acknowledges support from CNPq (Brazil) through a research productivity fellowship, grant no. [304873/2025-0]. S.K. acknowledges support from Science \& Technology Facilities Council (STFC)
(grant ST/Y001001/1).

This material is based upon work supported by the U.S. Department of Energy (DOE), Office of Science, Office of High-Energy Physics, under Contract No. DE–AC02–05CH11231, and by the National Energy Research Scientific Computing Center, a DOE Office of Science User Facility under the same contract. Additional support for DESI was provided by the U.S. National Science Foundation (NSF), Division of Astronomical Sciences under Contract No. AST-0950945 to the NSF’s National Optical-Infrared Astronomy Research Laboratory; the Science and Technology Facilities Council of the United Kingdom; the Gordon and Betty Moore Foundation; the Heising-Simons Foundation; the French Alternative Energies and Atomic Energy Commission (CEA); the National Council of Humanities, Science and Technology of Mexico (CONAHCYT); the Ministry of Science, Innovation and Universities of Spain (MICIU/AEI/10.13039/501100011033), and by the DESI Member Institutions: \url{https://www.desi.lbl.gov/collaborating-institutions}. Any opinions, findings, and conclusions or recommendations expressed in this material are those of the author(s) and do not necessarily reflect the views of the U. S. National Science Foundation, the U. S. Department of Energy, or any of the listed funding agencies.

The authors are honored to be permitted to conduct scientific research on I'oligam Du'ag (Kitt Peak), a mountain with particular significance to the Tohono O’odham Nation.

This research used resources of the National Energy Research Scientific Computing Center (NERSC), a Department of Energy User Facility (project desi-2025).

\end{acknowledgments}

\software{
Astropy              \citep{astropy1, astropy2, 2022Astropy},
Matplotlib           \citep{matplotlib},
NumPy                \citep{numpy},
pandas               \citep{pandas},
SciPy                \citep{scipy1, scipy2}}

\bibliography{references.bib}{}
\bibliographystyle{aasjournal}

\end{document}